\documentclass[twocolappendix]{aastex631}
\usepackage{amsmath}	
\usepackage{mathptmx}
\usepackage{graphicx} %use graph format  
\usepackage{epstopdf}
\usepackage{subfigure}
\usepackage{url}
\usepackage{makecell}
\usepackage{booktabs}
\usepackage{diagbox}
\usepackage{vcell}
\usepackage[T1]{fontenc}
\usepackage{amssymb}
\usepackage{color}
\usepackage{soul}

\usepackage{hyperref}

\shorttitle{SLT}
\shortauthors{Peng et al.}
\graphicspath{{./}{figures/}}
\begin{document}

\title[SLT]{Detection of Strongly Lensed Arcs in Galaxy Clusters with Transformers}

\correspondingauthor{Peng Jia, Nan Li and Hongyan Wei}
\email{robinmartin20@gmail.com}
\email{nan.li@nao.cas.cn}
\email{why\_1221@126.com}

\author[0000-0001-6623-0931]{Peng Jia}
\affiliation{College of Physics and Optoelectronic, Taiyuan University of Technology, Taiyuan, 030024, China}
\affiliation{Peng Cheng Lab, Shenzhen, 518066, China}
\affiliation{Department of Physics, Durham University, DH1 3LE, UK}

\author{Ruiqi Sun}
\affiliation{College of Physics and Optoelectronic, Taiyuan University of Technology, Taiyuan, 030024, China}

\author{Nan Li}
\affiliation{National Astronomical Observatories, Beijing, 100101,China}

\author{Yu Song}
\affiliation{College of Physics and Optoelectronic, Taiyuan University of Technology, Taiyuan, 030024, China}

\author{Runyu Ning}
\affiliation{College of Physics and Optoelectronic, Taiyuan University of Technology, Taiyuan, 030024, China}

\author{Hongyan Wei}
\affiliation{College of Physics and Optoelectronic, Taiyuan University of Technology, Taiyuan, 030024, China}

\author{Rui Luo}
\affiliation{Central South University, Changsha, 410083,China}



\begin{abstract}
Strong lensing in galaxy clusters probes properties of dense cores of dark matter halos in mass, studies the distant universe at flux levels and spatial resolutions otherwise unavailable, and constrains cosmological models independently. The next-generation large scale sky imaging surveys are expected to discover thousands of cluster-scale strong lenses, which would lead to unprecedented opportunities for applying cluster-scale strong lenses to solve astrophysical and cosmological problems. However, the large dataset challenges astronomers to identify and extract strong lensing signals, particularly strongly lensed arcs, because of their complexity and variety. Hence, we propose a framework to detect cluster-scale strongly lensed arcs, which contains a transformer-based detection algorithm and an image simulation algorithm. We embed prior information of strongly lensed arcs at cluster-scale into the training data through simulation and then train the detection algorithm with simulated images. We use the trained transformer to detect strongly lensed arcs from simulated and real data. Results show that our approach could achieve $99.63 \%$ accuracy rate, $90.32 \%$ recall rate, $85.37 \%$ precision rate and $0.23\%$ false positive rate in detection of strongly lensed arcs from simulated images and could detect almost all strongly lensed arcs in real observation images. Besides, with an interpretation method, we have shown that our method could identify important information embedded in simulated data. Next step, to test the reliability and usability of our approach, we will apply it to available observations (e.g., DESI Legacy Imaging Surveys \footnote{\url{https://www.legacysurvey.org/}}) and simulated data of upcoming large-scale sky surveys, such as the Euclid \footnote{\url{https://www.euclid-ec.org/}} and the CSST \footnote{\url{https://nao.cas.cn/csst/}}.
\end{abstract}



\section{Introduction} \label{sec:1}
Gravitational lensing has become a powerful probe in many areas of astrophysics and cosmology \citep{2012MNRAS.420..792M,2013ApJ...777...43M,2014RAA....14.1061F,2015IJMPD..2430020R,2017SchpJ..1232440B,2018ARAA..56..393M}. The phenomenon has been detected since \citet{1979Natur.279..381W} and over a wide range of scales, from Mpc in the weak-lensing regime \citep{2000MNRAS.318..625B,2003ApJ...597...98H,2005PhRvD..72b3516C,2008PhRvD..78d3002S,2016ApJ...817..179J,2017MNRAS.472.2126K,2018MNRAS.479.4998T} to kpc in strong lensing \citep{1986BAAS...18R1014L,1987Msngr..50....5S,1988A&A...200L..17F,1998ApJ...503..531H,1988ycs..work..156H,2002ApJ...571..712B,2017MNRAS.465.3185O,2018MNRAS.481L..40S,2018MNRAS.477..195T} and down to pc and sub-pc scales probed by microlensing \citep{2017MNRAS.467.1259B,2017ApJ...840L...3S,2018ApJ...867..136H}. Therefore, gravitational lensing can be used to measure the mass distribution in the Universe \citep{2013ApJ...765...24N,2015MNRAS.446.1356H,2018MNRAS.473.4279D,2018MNRAS.477.4046J}, improve the study of lensed high redshift galaxies \citep{2013ApJ...762...32C,2013ApJ...765...48J,2015MNRAS.453.4311S,2015MNRAS.452.2258D} and constrain cosmological parameters \citep{2013ApJ...766...70S,2014ApJ...788L..35S,2014ApJ...783...83L,2015ApJ...813...69M}, amongst other applications. Cosmological strong lensing is an extreme representation of gravitational lensing, which leads to multiple highly magnified and distorted images of background sources. In particular, highly distorted galaxies by strong lenses are called Gravitational lensed arcs, which have been used in various cosmological applications \citep{2011A&ARv..19...47K,2013SSRv..177...31M}. For instance, the frequency of strongly lensed arcs on the sky reflects the abundance \citep{2004ApJ...609...50D,2006ApJ...652...43L,2007A&A...473..715F,2010A&A...519A..91F,2007MNRAS.382.1494H}, the concentration \citep{2012MNRAS.420.3213O,2013MNRAS.434..878S,2014ApJ...797...34M} and astrophysical properties \citep{2008ApJ...687...22R} of massive lenses, and the redshift distribution and properties of sources \citep{2004ApJ...606L..93W,2011ApJ...727L..26B,2012ApJ...744..156B}. Expectedly, strongly lensed arcs can bring more reliable restraints on astrophysical and cosmological problems with the enormous data from next-generation surveys.\\

So far, the strongly lensed arcs have been detected almost exclusively by visual inspection of cluster images, although  automated search algorithms have recently been proposed \citep{2004A&A...416..391L,2005ApJ...633..768H,2006astro.ph..6757A,2007A&A...472..341S,2010MNRAS.406.1318H}. However, scanning wide-field data by eye covering hundreds (or even thousands) of square degrees for arcs appears as a hopeless endeavour. Regardless of the size of the datasets, detecting arcs with human inspection may involve potential biases due to seekers' capability. Some arcs may be obstructed by bright foreground galaxies or stars, which could not be detected by human experts. In contrast, automated tools for finding arcs bring objective and reproducible definitions of arc samples, especially in the case of blind searches on the large-scale sky across multiple survey projects \citep{2013SSRv..177...31M,2016ApJ...817...85X,2019MNRAS.482..313L}, as well as for the comparison between observational and simulated data \citep{2005ApJ...633..768H,2011MNRAS.418...54H,2019ApJ...878..122L}.\\

More recently, machine learning algorithms have become the mainstream for identifying gravitational lenses automatically \citep{2018MNRAS.477.2841M}, given their strong performance in the field of general image recognition. In particular,  detecting the galaxy scale strong lensing systems has reached a remarkable success \citep{2017MNRAS.471..167J,2017MNRAS.472.1129P,2017MNRAS.465.4325O,2017A&A...597A.135B,2017MNRAS.471.3378H,2019ApJ...877...58A,2018MNRAS.473.3895L}, and the number of high-quality candidates of strong lensing systems is over a couple of thousands by combining the data products of DES, DESI, and KIDS \citep{2019yCat..18630089S,2020ApJ...899...30L,2020ApJ...894...78H,2021ApJ...909...27H, 2019MNRAS.482..313L}. Notably, most above studies are based on stamps centred at galaxies, which is reasonable for galaxy scale lenses but unsuitable for strong lensing in galaxy clusters (CGSLs), for the following three reasons:\\
 1. Since the number of CGSLs is too small, it is not possible to solely use these known CGSLs as the training set.\\
 2. Since not all CGSLs are around the centres of galaxy clusters \citep{Meneghetti2020Sci}, we would loss lots of CGSLs which does not have Brightest Cluster Galaxies (BCGs), if we use BCGs as prior condition for detection of CGSLs.\\
 3. Since CGSLs are rare and may extend to very large scale, if we split full frame images to patches of images with smaller size for classification, we would obtain a lot of false positive detection results, even when the false positive rate is low.\\
Therefore, we propose a framework to detect CGSLs, which contains two parts: a detection algorithm and a simulation algorithm. The simulation algorithm would embed prior information of CGSLs known by scientists into images in the training data set. After training with simulated images, the detection algorithm could detect CGSLs from full frame observation images across a large field of view without the centring and cutting-out process. With these two algorithms, we could detect CGSLs according to prior information provided by scientists.\\

For the image simulation part, we assume arcs are features of CGSLs and we could use these features to detect CGSLs. We use the PICS \citep{li2016pics} to generate ideal images that contain CGSLs according to extragalactic catalogue CosmosDC2 \citep{2019Korytov} and their corresponding labels (mask matrix with the same size of simulated images) in the training set. Besides, we have also generated images that does not contain CGSLs and zero labels (zero matrix with the same size of simulated images) in the training set. Although the percentage of CGSLs in real observation images is very small, we will still set half of total images contain CGSLs to increase the training efficiency. For simulated images in the validation set, we would also set half of total image contain CGSLs and the rest part of images does not contain CGSLs to evaluate the performance of our algorithm. Besides, we would also generate simulated images and $1\%$ of them contain CGSLs as a test set to test the performance of our algorithm in real applications.\\
 
For the detection algorithm, the Convolutional Neural Network (CNN) is widely used as the basic structure. However, the performance of the CNN based detection algorithm is limited by the receptive field of the convolutional kernel. For targets with variable scales and complex structures, the performance of CNN based detection algorithms would be limited. In our previous paper, we have found that for classification of simple point--like or streak--like astronomical targets observed by wide field small aperture telescopes, a trained recurrent neural network (RNN) could have better performance than that of the CNN \citep{jia2019optical}. The RNN has a sequence structure. After training, the RNN may better capture features with larger size. However, the RNN treats the whole image as a long sequence, which would require a lot of GPU memory and cost long time during the training stage. Thanks to \citet{vaswani2017attention}, an attention based neural network, transformer has been proposed. The transformer does not need sequence-aligned recurrent architecture, which makes it easier to train, even with large number of parameters. The transformer is firstly used for natural language processing and then for image processing. For target detection tasks, the DEtection TRansformer (DETR) is widely studied since it was proposed by \citet{carion2020end}. The DETR uses transformer to reason about the relations between objects to be detected and the global image context to directly predict positions and types of targets. The mechanism of the DETR is similar to that of human attention, which would extract semantic information of images for detection and could achieve better performance in detection of complex and extended targets.\\

CGSLs are a type of celestial objects with complex and extended shapes, which contain some front galaxies in the centre and some arc structures around the centre. These features could be used as semantic information for attention-based source detection algorithms. In the algorithm developed by \citet{thuruthipilly2021finding}, features from attention based encoder layers are extracted for classification of SGLs at galaxy scale from candidate images and have achieved better performance than that of the CNN based methods. Since CGSLs have much larger size and more irregular shape than SGLs at galaxy scale, we propose to integrate the DETR and the Deformable DETR \citep{zhu2020deformable} with ensemble learning strategy to build detection algorithm with better performance. Because detection results would be checked by human scientists for further study, the true positive rate should be high and the recall rate should be moderate. Therefore, we further propose a two-step strategy for detection of CGSLs to increase its performance in real applications.\\

For the validation set with half of total images contain CGSLs, our method could achieve more than $88 \%$ recall rate and more than $70\%$ precision rate, when we directly use our method with IOU of 0.1 and score of 0.7. Considering many CGSLs contain BCGs and they could not be detected even by human inspections, the recall rate and the precision rate is acceptable. For the test set with $1\%$ of total images contain CGSLs, our two step detection strategy could achieve $99.63 \%$ accuracy and $0.23 \%$ false positive rate, when the recall is $90.32\%$ and the precision is $85.37 \%$. We further use real observation images from the Hubble Space Telescope Frontier Fields project, the Hubble Space Telescope RELICS project and the early release image from the James Webb Space Telescope to detect CGSLs. For real observation images, we find that almost all CGSLs could be detected by our algorithm, except several false detection results brought by diffraction rings in these images, which are not included in the training data. At last, we use the interpretation method to show that our detection algorithm could focus on important features (arcs) of CGSLs.\\
 
In Section \ref{sec:2}, we firstly describe the simulation and data processing method to obtain training data for our algorithm. Then we will analyse the detection requirements of CGSLs and adopt our evaluation criterion for the detection algorithm. The basic structure and the performance of the DETR are shown in Section \ref{sec:3}. Section \ref{sec:4} describes a comparative investigation between DETR and Deformable DETR, including training and detection performance. In Section \ref{sec:5}, we explore the correlation between the features of images and the detection performance with a machine-learning-interpretation module. In Section \ref{sec:6}, we will show the performance of our algorithm in deploying our algorithm with simulated and real observation data.  At last, discussions and conclusions are delivered in Section \ref{sec:con}.\\

\section{Data Preparation Procedure and Evaluation Criterion for Detection of CGSLs} \label{sec:2}
As we have discussed in Section \ref{sec:1}, only tens of  CGSLs are discovered so far. The number is too small to be used as the training set. Besides, many CGSLs are discovered by the visual system of human beings, which would introduce statistical bias into training data. A neural network trained with these data may only be possible to find `similar' CGSLs. Besides, considering detection of CGSLs from images of multiple bands is beyond the capacity of human vision systems, a lot of CGSLs would be lost by a detection neural network that are trained by data obtained by human vision systems. Therefore, we use simulation data to train the neural network, which could embed prior information about scientists' understanding of CGSLs into images in the training set. Meanwhile, we could also enlarge the volume as well as the diversity of the training data with the simulation algorithm. After training, the detection algorithm would be able to detect CGSLs that satisfy scientists' prior assumption about properties of CGSLs and are missed by human vision inspections.\\

We will briefly introduce the simulation method in Subsection \ref{subsec:21} and introduce data processing strategy to generate training data in Subsection \ref{subsec:22}. Besides, although the mean average precision (mAP) is widely used as evaluation criterion for general target detection algorithms, for CGSL detection tasks, we will show that it would be better to use Precision and Recall under a predefined intersection over union (IOU) as the evaluation criterion. In Subsection \ref{subsec:23}, we will discuss the evaluation criterion for the CGSL detection algorithm.\\

\subsection{Simulation of CGSLs}
\label{subsec:21}
To train and evaluate neural network models for detection of CGSLs, we have created an ideal synthetic dataset without PSFs and noises using a simulation pipeline named PICS \citep{li2016pics}. Similar to \citet{Madireddy2019}, the simulation of CGSLs in this paper comprises six steps: (1) create populations of lenses and sources according to the given statistical properties of CGSLs; (2) build mass and light models of foreground lenses; (3) calculate deflection fields of the lenses; (4) construct light profiles of background source galaxies; (5) run ray-tracing simulations to create strongly lensed images based on the deflection fields and light profile of sources; and (6) stack the lensed images of lensed arcs and images of galaxies on the line-of-sight as well as the foreground images of lenses.\\

The populations of lenses and sources are built based on a state-of-the-art extragalactic catalogue called CosmoDC2 \citep{2019Korytov}. CosmoDC2 provides a catalogue of galaxy clusters, including the virial mass of dark matter halos and the apparent magnitudes, axis ratios, position angles, and redshifts of member galaxies. The mass model of a lens galaxy cluster is modelled as a dark matter halo plus a set of member galaxies. The mass model of dark matter halo is Elliptical NFW, and it requests virial mass, concentration parameter, and ellipticity. CosmoDC2 gives the virial mass, concentrations are calculated according to the c-M relation given by \cite{Child2018}, and ellipticity is obtained by measuring the ellipticity of the spacial distribution of member galaxies in the cluster. Hence, the deflection angle map due to dark matter halo can be described by ${M_{vir}, c_{vir}, q_{nfw}, z_l, z_s}$, where $z_l$ and $z_s$ are the redshifts of lens plane and source plane separately.\\

The mass model of member galaxies is a singular isothermal ellipsoid (SIE) as adopted in \cite{Collett2015}, since SIE is analytically tractable and consistent with models of individual lenses and lens statistics on length scales relevant for strong lensing \citep{Koopmans2006, Gavazzi2007, Dye2008}. Accordingly, the deflection maps due to member galaxies can be defined by positions, velocity dispersions, axis ratios, position angles, and redshifts of member galaxies as well as redshifts of source galaxies, namely, ${x_1, x_2, \sigma_v, q_l, \phi_l, z_l, z_s}$. The parameters ${x_1, x_2, q_l, \phi_l, z_l, z_s}$ are taken directly from the CosmoDC2 catalogue. $\sigma_v$ is derived from the $L-\sigma$ scaling relation from the bright sample of \citet{parker2007masses} given by
\begin{equation}
\sigma_v=142 (L/L_{star})^{1/3l} kms^{-1},
\end{equation}
where $log10(L/L_{star}) = -0.4(magr - magr_{\star})$, and magr is the apparent r-band magnitude of the galaxy given by the CosmoDC2 catalogue. We adopt the assumption in \cite{More2016} that $magr_{\star}$ evolves with redshift as $magr_{\star} = + 1.5(z - 0.1) - 20.44$ \citep{Faber2007}. To guarantee significant lensing features, we set $z_s > z_l+0.5$, then randomly choose galaxies satisfying this criterion. The projected positions of sources in the lensing system are randomly chosen in the area where lensing magnifications are larger than 20 on the source plane. The light profile of galaxies in the light cone are all modelled as a composite Sersic profile containing bulges and disks, and all the parameters are from the CosmoDC2. For non-lensed galaxies, including member galaxies and line of sight galaxies, the images are rendered with composite Sersic profiles on regular grids directly; for lensed arcs, the images are rendered with composite Sersic profiles on ray-traced grids.\\

The final simulated data are generated by stacking images of background galaxies and images of generated CGSLs. We also provide corresponding masks of lensed arcs in image stamps to generate labels for the training data. Each simulated image includes four channels. The first three channels are images of g, r and i bands and the last channel is the mask of lensed arcs. The size of simulated images is $1280 \times 1280$ pixels and CGSLs are majorly in the centre of these simulated images. Besides, we have also generated images that contain galaxies without CGSLs. For these images, the first three channels are images of g, r and i band and the last channel is a zero matrix with the same size of images in the first three channels. Some additional data processing are required to generate training data from these simulated images, which we will discuss in Subsection \ref{subsec:22}.\\ 

\subsection{Generation of Training Data from Simulated CGSL Images}
\label{subsec:22}
\begin{figure*}
\centering
\subfigure[]{
\includegraphics[width=0.31\textwidth]{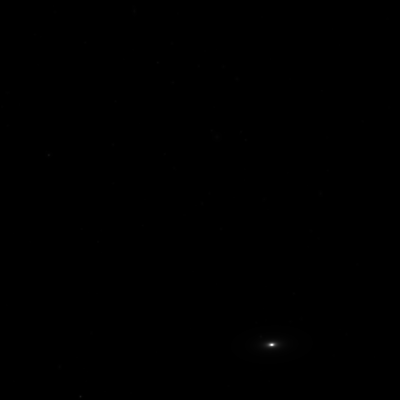}
}
\quad
\subfigure[]{
\includegraphics[width=0.31\textwidth]{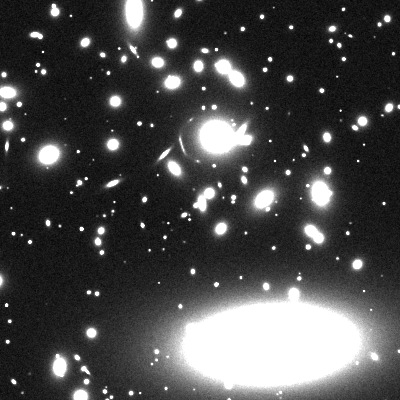}
}
\quad
\subfigure[]{
\includegraphics[width=0.31\textwidth]{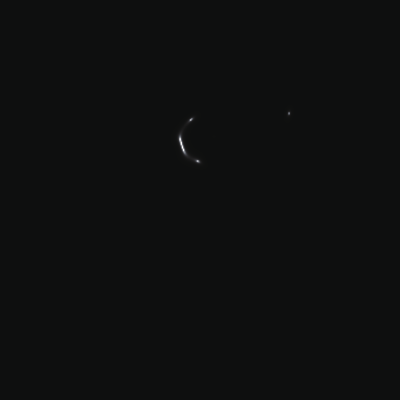}
}
	\caption{This figure shows the original simulated image in figure 1.a, the image after greyscale transformation in figure 1.b and the mask of lensed arcs in figure 1.c. As we can see, structure of this image is almost invisible in the original image. After greyscale transformation, we can easily see various galaxy structures and even directly observe the CGSL, which would be beneficial to development of the algorithm. We can see that the mask of lensed arcs ignores the foreground and shows the boundary of the CGSL, which is an arc in this image.}  
	\label{figure1}  
\end{figure*}

The detection method proposed in this paper is a supervised learning algorithm. For a supervised learning algorithm, training data should include both its inputs (images of CGSLs) and its outputs/labels (positions and size of CGSLs). We firstly need to generate labels and images from simulated data. Contemporary detection algorithms use bounding boxes to indicate positions and sizes of targets and bounding boxes are circumscribed rectangles of targets. However, masks of CGSLs are two dimensional mask images with the same size as that of simulated images. Therefore, we would transform greyscale values of masks of lensed arcs with log transformation. Then pixels with grey scale values larger than $10^{-3}$ will be set as part of targets and other pixels in the mask images will directly be set as backgrounds, as shown in figure~\ref{figure1}. Then we would generate circumscribed rectangles for pixels belong to CGSLs as bounding boxes.\\

For inputs of CGSL detection algorithms, we have noticed that difference between grey scale values in simulated images is large and it is hard to see features of CGSLs directly from simulated images. Generally greyscale transformation such as zscale in DS9 \citep{2000ascl.soft03002S} would make structures of CGSLs easier to be detected. Therefore, we would transform grey scales of simulated images with zscale transformation. Although neural networks could learn the grey scale transformation algorithm, it would be easier and faster to train neural networks, if we have made grey scale transformation before training.\\

\begin{figure*}
\centering
\subfigure[]{
\includegraphics[width=0.15\textwidth]{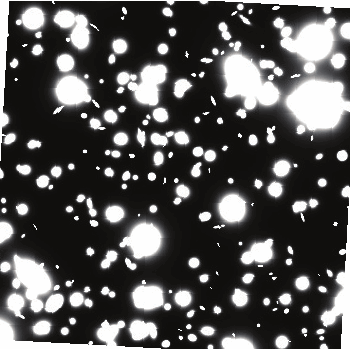}
}
\quad
\subfigure[]{
\includegraphics[width=0.15\textwidth]{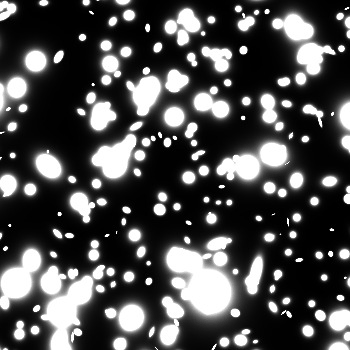}
}
\subfigure[]{
\includegraphics[width=0.15\textwidth]{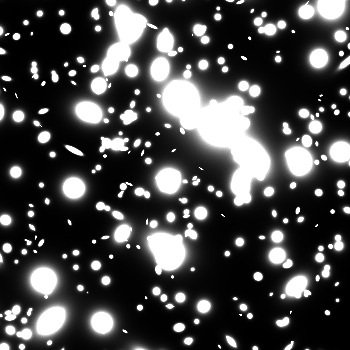}
}
\quad
\subfigure[]{
\includegraphics[width=0.15\textwidth]{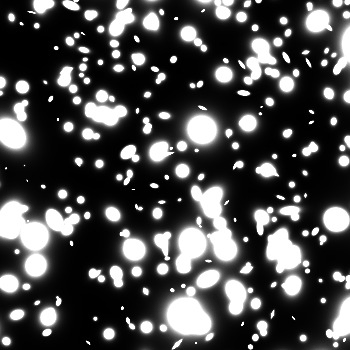}
}
\subfigure[]{
\includegraphics[width=0.15\textwidth]{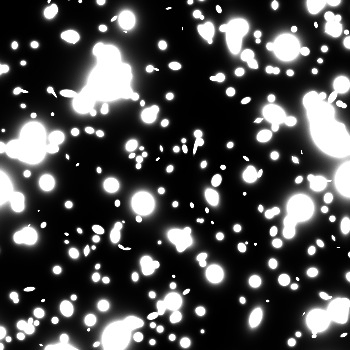}
}\\
\quad
\subfigure[]{
\includegraphics[width=0.15\textwidth]{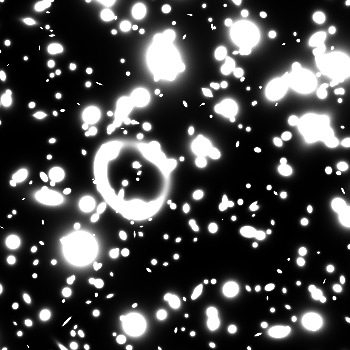}
}
\quad
\subfigure[]{
\includegraphics[width=0.15\textwidth]{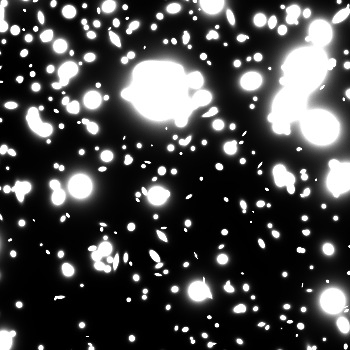}
}
\subfigure[]{
\includegraphics[width=0.15\textwidth]{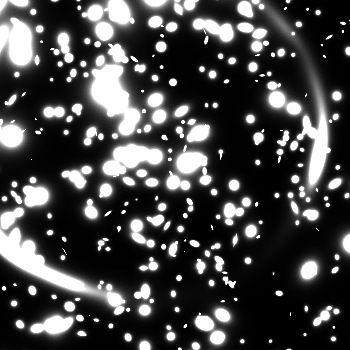}
}
\quad
\subfigure[]{
\includegraphics[width=0.15\textwidth]{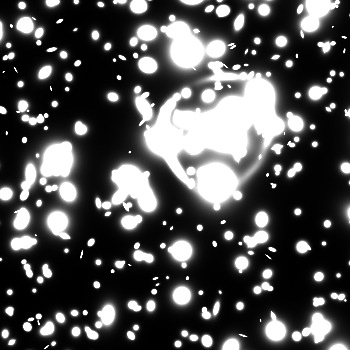}
}
\subfigure[]{
\includegraphics[width=0.15\textwidth]{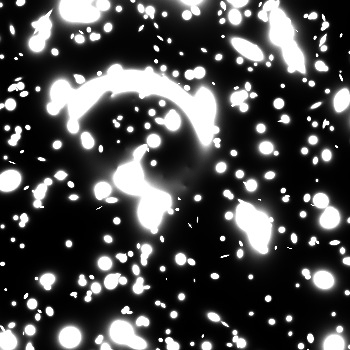}
}
\caption{This figure shows several images without and with CGSLs. Figure \ref{figure0}.a \ref{figure0}.b \ref{figure0}.c \ref{figure0}.d \ref{figure0}.e show images without CGSLs and figure  \ref{figure0}.f \ref{figure0}.g \ref{figure0}.h \ref{figure0}.i \ref{figure0}.j show show images with CGSLs. We can find that in the images with CGSLs there are arcs, but in the images without CGSLs there are no similar structures. }
\label{figure0}
\end{figure*}

After grey scale transformation, images of the same target in different bands will be stored as an image with several channels. Besides, since almost all strong lensing targets are in the centre of simulated images, it would introduce strong bias into the detection algorithm during the raining stage. The detection algorithm would be more likely to predict positions of CGSLs in the centre of these images. Therefore, we generate a window of $400 \times 400$ pixels and randomly shift the window in the original images to cut stamp images for detection. With this method, there are some images without any CGSLs or only with part of CGSLs. Besides, many CGSLs would distribute in different parts of these stamp images. Figure ~\ref{figure0} shows several images with and without CGSLs. It should be noted that we have generated images with a size of $400 \times 400$ pixels as the training set to reduce the requirement of GPU memory during the training stage. Since the DETR and the Deformable DETR could accept images of any size as inputs, we would directly detect CGSLs from full frame images, as long as we have enough GPU memory during the deployment stage. At last, all images will be saved as PNG files with 3 channels as inputs of the neural network, as shown in figure~\ref{figure0}. It should be noted that difference of grey scale values in some pixels of original images could be quite large, which would introduce difficulties in development of detection algorithms. Transforming original simulated images to PNG files, whose grey scale values are integers within the scale of 0 to 255, can constrain data distribution and make our algorithm easier to train at the cost of low detection ability of dim targets and higher position regression error.\\

\subsection{Performance Evaluation Criterion for Detection of CGSLs}
\label{subsec:23}
The performance evaluation criterion of the detection algorithm is important for the algorithm development. Although the mean average precision (mAP) is widely used to evaluate the performance of general purpose detection algorithm, it is better to use an appropriate evaluation criterion, according to real detection requirements. Hence, we select the precision rate and the recall rate under a predefined Intersection over Union (IOU) ratio as the performance evaluation criterion. The IOU is defined as the ratio between the over--lap area and the union area of the bounding box from the detection results and the labels as defined in equation ~\ref{equation1}.\\
\begin{equation}
IOU = \frac{Intersection Area}{Union Area}
\label{equation1}
\end{equation}

If the IOU is larger than a predefined criterion, we set the detection as a true positive detection (TP). Otherwise, we set the detection result as a false negative (FN) or a false positive (FP) detection. Then we would further define Precision and Recall rate to evaluate the performance of a detection algorithm with TP, FP and FN. The Precision rate is the percentage of true positive detection results to all detection results and the recall rate is the percentage of true positive detection results to all targets. The Precision and the Recall rates are defined in equation ~\ref{equation2}.\\
\begin{equation}
\begin{split}
Precision = \frac{TP}{TP+FP}\\
Recall = \frac{TP}{TP+FN}
\label{equation2}
\end{split}
\end{equation}

Because CGSLs are very rare, normally less than one in a full frame observation image. Therefore it would be more important to discover one CGSL instead of getting the accurate position of the CGSL in observation images. Therefore, we could firstly set a small value of IOU to define the Precision and Recall rate, such as 0.1 in this paper. If we set the IOU to be 0.1, the overlap between the detection result (bounding box) and the true result (true position and size of the CGSL) would be larger than $10\%$ of the true result. Therefore, the detection result would roughly indicate the position of the CGSL. Then we need to further use the segmentation algorithm to obtain CGSLs from detection results. For the image segmentation algorithm, we need to expand the size of the bounding box by 10 times to make sure the whole image of the CGSL is in the bounding box.\\

\section{Detection of CGSLs with the DETR} \label{sec:3}
As mentioned in Section \ref{sec:1}, CGSLs have extended and complex structures and would often be obstructed by BCGs, as shown in Figure~\ref{figure0}.g, Figure~\ref{figure6}.a, Figure~\ref{figure12}.g and Figure~\ref{figure12}.h. In this Section, we would describe our strategy to use the DETR to detect CGSLs. We will give a brief introduction of the DETR in Subsection \ref{sec:31}. We will further describe the training strategy and show the performance of the DETR in detection CGSLs in Subsection \ref{sec:32}.\\ 

\subsection{Introduction of the DETR}
\label{sec:31}
\begin{figure*}
	\centering  
	\includegraphics[width=\textwidth]{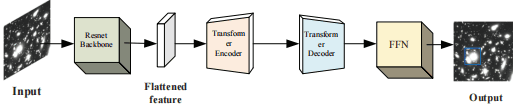}  
	\caption{The data flow-chart of the DETR. It includes an encoder and decoder part which could learn features of extended targets for detection.}  
	\label{figure2}  
\end{figure*}

The data flow-chart of the DETR is shown in Figure~\ref{figure2}. When we input an image into the DETR, its features of different pixels will firstly be extracted by the back-bone neural network to form different feature vectors. It should be noted that since these feature vectors are related to different pixels in the image, feature vectors of adjacent pixels would have similarities. However, feature vectors are abstract representations of contents by the back-bone neural network, so it does not include the corresponding position information. Therefore, feature vectors and their corresponding position vectors will be connected together as one dimensional vector and sent to the transformer. In the transformer, these feature vectors with position information would be processed again by means of the attention mechanism. At last, results will be sent to a feed-forward propagation (FFP) neural network for target position and type prediction.\\

\begin{figure*}
	\centering  
	\includegraphics[width=\textwidth]{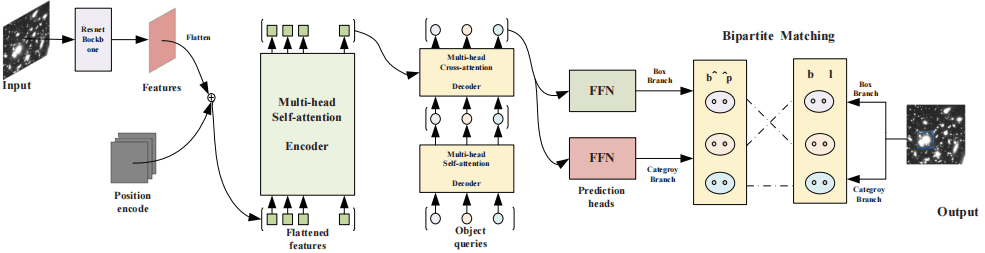}  
	\caption{The structure of the DETR. The input image is firstly extracted by the Backbone neural network to obtain features. After adding the encoded position information of feature vectors, feature vectors are sent to an encoder-decoder structure. The output decoded vectors will get the category and the position information of the target in the input image through two FFN branches respectively. The lastbipartite-matching structure of the network is to match the prediction output with the label of the input to help the network to achieve the correct prediction result. The variables $\hat{p}$ and $\hat{b}$ respectively represent the category and position box predicted by the algorithm, $l$ and $b$ represent the real category and position box of the target, and lines represent the process of cross matching between the predicted results and real labels of this bipartite-matching part. Each oval is an object, which could either be the background or a target. The circles inside the ovals is the value of position or category. }  
	\label{figure3}  
\end{figure*}

The structure of the DETR used in this paper is shown in Figure~\ref{figure3}. We propose to use the Resnet50 as the backbone neural network for feature extraction, because the Resnet50 has a residual structure, which makes it easier to train \citep{he2016deep}. Besides, the Resnet50 could gain high accuracy even with very deep structure. The feature vectors extracted from backbone neural network would be sent to the transformer along with the position information of these feature vectors. The position vectors which represent position information of these feature vectors are encoded by equation~\ref{equation3}:\\
\begin{equation}
\begin{split}
PE(x,y)_{2i} = \sin\left ( (x,y)/1000^{2i/d_{model}}\right )\\
PE(x,y)_{2i+1} = \cos\left ( (x,y)/1000^{2i/d_{model}}\right )
\label{equation3}
\end{split}
\end{equation}
where $(x,y)\in [0,1]^{2}$ is the normalized position of the current feature vector in the image and $d_{model}$ is the dimension of position vectors, which has the same size as that of the feature vector or the dimension of the hidden state of the DETR. $i$ represents the location index of each small piece at different position in the current feature vector. We encode these pieces in even position and odd position with sin and cos functions respectively, and finally obtain the position vector $PE(x,y)$ of the feature vector at $(x,y)$ with $d_{model}$ dimensions. At last, we would directly add the position information into the corresponding feature vector.\\

\begin{figure*}
	\centering  
	\includegraphics[width=\textwidth]{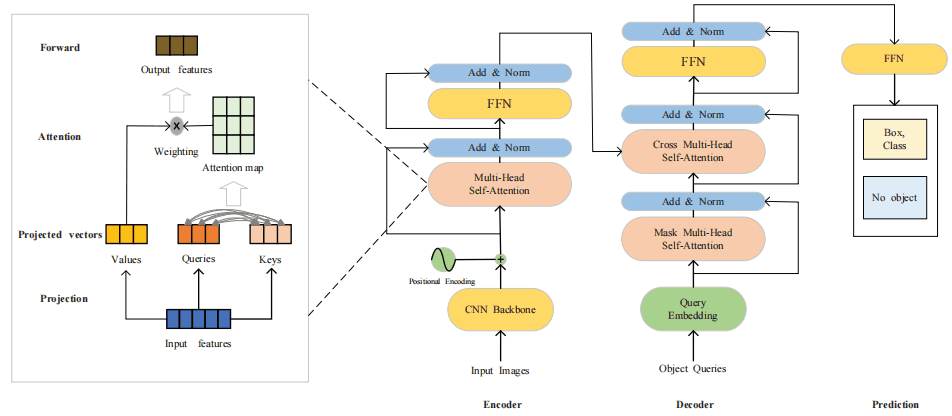}  
	\caption{The encoder-decoder structure of the transformer. Features of images are firstly extracted by the CNN backbone and then the multi-head self-attention layer would generate attentions of these features in the encoder. In the decoder, several Object Queries will sample features automatically and the Mask Multi-Head Self-Attention Layer will process these features with self-attention mechanism. At last, in the Multi-Head Cross-Attention layer, features will further be processed to generate cross-attention to detect targets.}  
	\label{figure4}  
\end{figure*}

The transformer has an encoder-decoder structure as shown in Figure~\ref{figure4}. The encoder part is designed based on the attention mechanism and it is called self-attention layer. If a feature vector is put into the self-attention layer, it would be transformed to three vectors: Value, Query and Key. Similar to hidden states in RNN, Value, Query and Key are used to project feature vectors to different directions and positions and obtain relations between features. For each feature vector, its Query will multiply Keys of all feature vectors to calculate the correlation between the current vector and all other feature vectors. Therefore for all feature vectors, we could get a weighted map which is called attention map and values in each pixel of the weighted map stand for correlations between features. Higher correlation values between features mean they have attracted more attention and vice versa. The attention map will propagate to a softmax layer and outputs of the softmax layer will be multiplied by values of all feature vectors to get final outputs. The self-attention layer would naturally obtain connections between features, regardless of their distance, which could exceed the receptive field of convolutional kernels in CNN. Several attention layers will be connected together to generate a head. We would use several heads to design our detection neural network, based on the multi-head mechanism. With the multi-head design, we could extract more information to focus on different positions in an image.\\

The decoder is also designed with the concept of attention. Inputs of the decoder are random initialized vectors which are called Object Queries. Object Queries are similar to ``anchors'' in the Faster-RCNN \citep{2015arXiv150601497R}. Each object query represents a target predicted by the transformer based detection algorithm. In other words, the number of object queries is the number of objects predicted by the model, including the prediction results of background (no object). The number of queries is fixed and would be much larger than the actual number of targets in the input image. Prediction results would be sent to the Feed Forward Network (FFN). In the FFN, the predictions of the model will be bipartite-matched with input labels, where a majority of queries will be classified to the background or ``no object'' and only a few closest to the actual target will be regarded as predictions of real targets. During the training stage, Object Queries will automatically sample features of objects from the training set and propagate these features to Mask Multi-Head Self-Attention layer. In the Mask Multi-Head Self-Attention layer, features will be processed again by the self-attention mechanism. Then outputs of Mask Multi-Head Self-Attention layer and outputs from the encoder will be sent to the Multi-Head Cross-Attention layer. The multi-head cross attention layer is designed according to the cross attention mechanism. There are also three vectors in the cross-attention mechanism: Value, Query and Key and we would carry out the same operation process as the self-attention mechanism. The difference is that the Query comes from decoder, while the Key and the Value come from encoder. The output from the encoder and that from the decoder is crossed, so it is called the cross attention mechanism. In the Multi-Head Cross-Attention layer, features will further be processed to generate cross-attention to better focus on interested targets.\\

At last, output feature vectors will be sent to two FFNs for object classification and position regression. Different from other CNN based detection algorithms, we would use outputs from FFN and labels to calculate loss of the DETR. Because the total number of predictions is fixed in the DETR and more predictions would require larger memory, so the DETR is better in detection of sparsely distributed large astronomical targets, such as CGSLs. With N predictions, the Hungarian algorithm is used in the last bipartite-matching structure to match N outputs from the FFN layer and M labels, resulting in the Hungarian loss between the predictions $\hat{y}$ and the label $y$. The Hungarian loss is a combinatorial algorithm that are used to calculate optimal matching between prediction results and ground truth \citep{kuhn1955hungarian}, which is defined in equation \ref{equation4}, \\
\begin{equation}
\mathit{L}_{\textsl{Hungarian}}(y,\hat{y})=\sum_{i=1}^{N}\left [ \mathit{L} _{\textsl{class}}(i,\hat{\sigma (i)}) + \mathit{L}_{box}(i,\hat{\sigma (i)})\right].\\
		\label{equation4}
\end{equation}
There are two types of losses in the Hungarian loss: the matching classification loss $L_{class}$ and the bounding box loss $L_{box}$. For CGSL detection tasks, outputs could be classified either to be CGSL or background. Therefore, the classification loss is defined as cross-entropy between these two classes. The bounding box loss is used to evaluate position accuracy of the DETR outputs. $L_1$ loss and $L_{iou}$ loss are both used as defined in equation \ref{equation5},\\

\begin{equation}
\begin{split}
\mathit{L}_{\textsl{class}}(i,\hat{\sigma (i)})=\sum_{i=1}^{N}-(c_{i}*log(p_{\hat{\sigma (i)}})+(1-c_{i})*log(1-p_{\hat{\sigma (i)}})),\\
\mathit{L}_{\textsl{box}}(i,\hat{\sigma (i)})=\sum_{i=1}^{N}(\lambda _{iou}\mathit{L}_{iou}(b_{i},b_{\hat{\sigma (i)}})+\lambda _{L_{1}}||b_{i}-b_{\hat{\sigma (i)}}||_{1}),
\label{equation5}
\end{split}
\end{equation}

\begin{equation}
\begin{split}
\mathit{L}_{iou}(b_{i},b_{\hat{\sigma (i)}})=1-IOU
\label{equation7}
\end{split}
\end{equation}
where $i$ is the index of the target, and $\hat{\sigma (i)}$ is the corresponding prediction index, $c_{i}$ is the value of the target's category, $p_{\hat{\sigma (i)}}$ is the value of the prediction's category. $\lambda _{iou}$ and $\lambda _{L_{1}}$ are hyper--parameters corresponding to weights of $\mathit{L}_{iou}$ and $L_{1}$ norm respectively. $b_{i}$ is the position of the target and $b_{\hat{\sigma (i)}}$ is the predicted position. $L_{1}$ loss is the absolute error between the predicted position box and the position of real target box. In real applications, we find that targets of different size have different sensitivity to the $L_{1}$ loss. A larger target has a larger boundary box, so a small deviation will cause a large $L_{1}$ loss, while a smaller coordinate box may not cause a large $L_{1}$ loss, even if the deviation is large. To balance this, we introduce the box $\mathit{L}_{iou}$ loss which is independent of the size of the bounding box. The $\mathit{L}_{iou}$ loss is defined in equation \ref{equation7} and the IOU is defined in equation \ref{equation1}.\\
 
\subsection{Training and Performance Evaluation of the DETR}
\label{sec:32}
\begin{figure*}
\centering
\subfigure[]{
\includegraphics[width=0.46\textwidth]{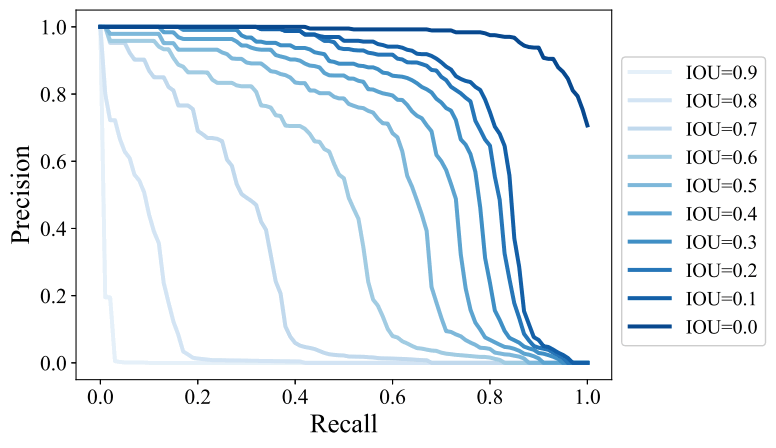}
}
\quad
\subfigure[]{
\includegraphics[width=0.46\textwidth]{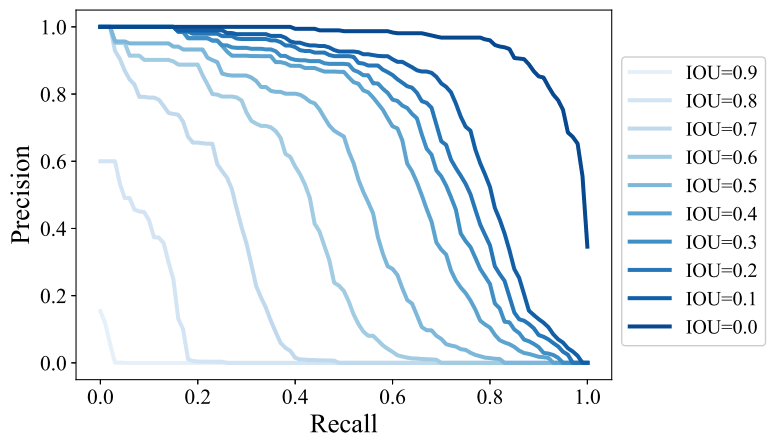}
}
\caption{The performance of the DETR and the Deformable DETR in detection of CGSLs. The vertical axis represents Precision Rate and the horizontal axis represents Recall Rate. The farther to the right of the curves, the higher the recall, and the higher the curves, the better the precision. We can use area of the P-R curve to evaluate the performance. The larger the area surrounded by the curves, the better the overall performance of the algorithm. IOU represents the degree of overlap between the predicted targets and the real targets. The larger IOU is, the greater the overlap is, indicating that the location of the targets predicted by the algorithm is more accurate. We take different IOUs as the threshold, and only when the overlap degree between the targets predicted by the algorithm and the real targets is larger than the threshold, we consider the prediction to be accurate.}
\label{figure5}
\end{figure*}

Because the number of CGSLs is small and the DETR has relatively complex structure, it would be hard to train the DETR. Therefore, we propose the following strategies to train the model:\\
1. We propose to add the drop-out strategy with drop-out rate of 0.1 to prevent over-fitting.\\
2. We use instance normalization to process each channel of an image (subtract average value of each channel and then divide standard deviation of these images).\\
3. We propose to use pre-trained weights to initialize the DETR to reduce training time. The pre--trained weights used in this paper is obtained from \citet{carion2020end}.\\
The DETR is implemented with the Pytorch \citep{paszke2019pytorch} in a computer with one Nvidia RTX 3090 GPU card. With the strategies proposed above, we train the DETR with Adams algorithm as the optimizer \citep{kingma2014adam}. There are two thousand images in the training set and six hundred images in the  validation set. We set half of images contain CGSLs and half of them does not contain CGSLs to help us to better evaluate performance of our algorithm. The DETR is trained with 250 epochs and costs around 25 hours. After training, we use the evaluation criterion defined in Subsection \ref{subsec:23} to evaluate the performance of the DETR in detection of CGSLs. It would cost about 0.06 seconds to process an image with size of $400\times 400$ pixels. The performance of our algorithm in detection of CGSL is shown in the left panel of figure~\ref{figure5}. It can be seen that the area surrounded by the P-R curve increases rapidly as the IOU threshold decreases, which indicates that the performance of our detection algorithm improves rapidly if we reduce position accuracy requirements, which is consistent with our discussions in Section \ref{subsec:23}.\\

If we set higher IOU, position accuracy of the detection algorithm would be higher at the cost of low recall rate. Considering that the number of CGSLs is small in real applications, we set the IOU threshold to be 0.1, which will enlarge the size of detection results by 10 times for further analyse. The area of the P-R curve can reach 0.85, when the IOU threshold is 0.1. Besides, it is worth noting that, when our IOU threshold is set as 0, the detection algorithm becomes a classification algorithm, which means that our algorithm can directly identify whether there is a strong gravitational lensing system in the image regardless of its position. For the classification task, we can see that the area of the P-R curve could reach around 0.95.\\

\begin{figure}
\centering
\subfigure[]{
\includegraphics[width=0.2\textwidth]{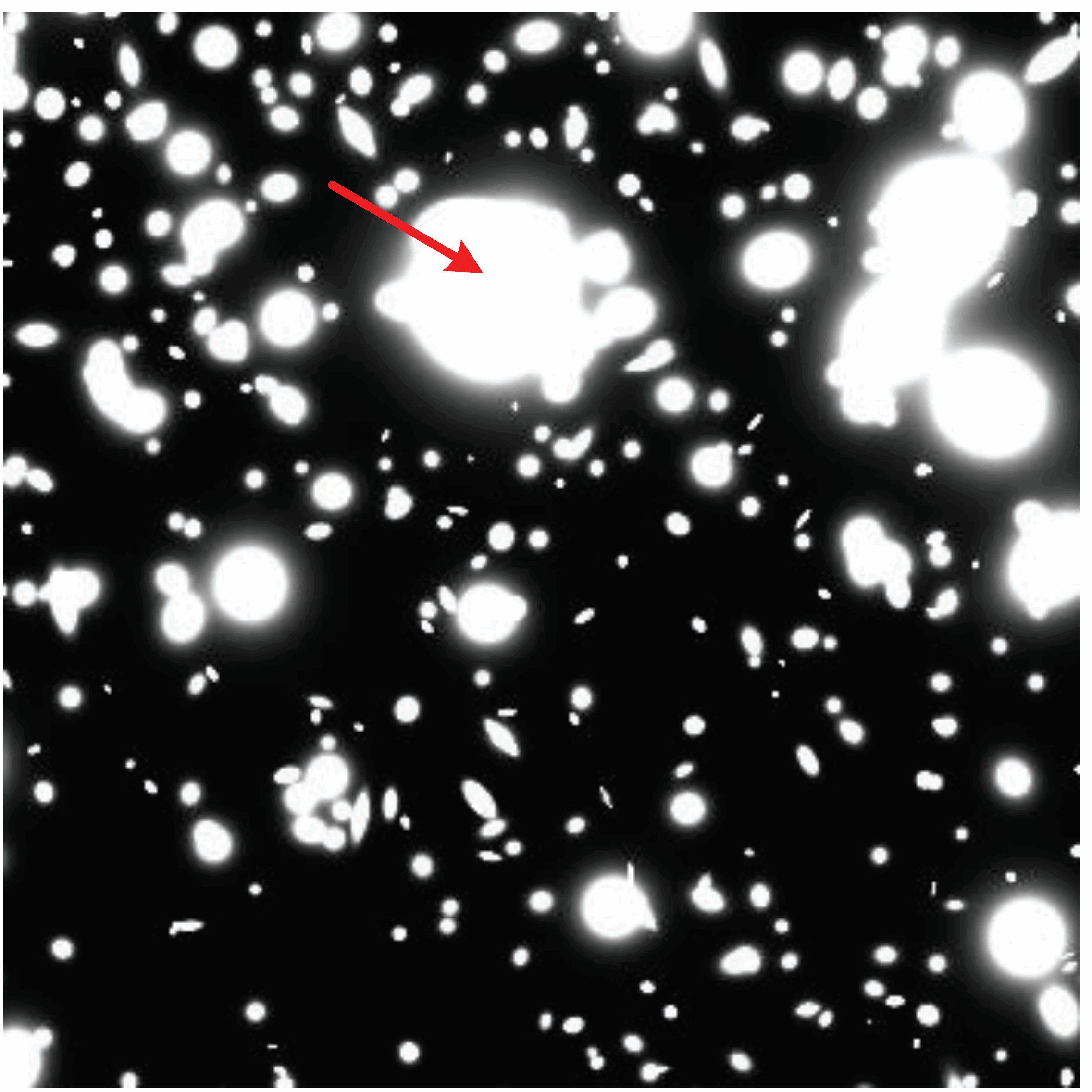}
}
\quad
\subfigure[]{
\includegraphics[width=0.2\textwidth]{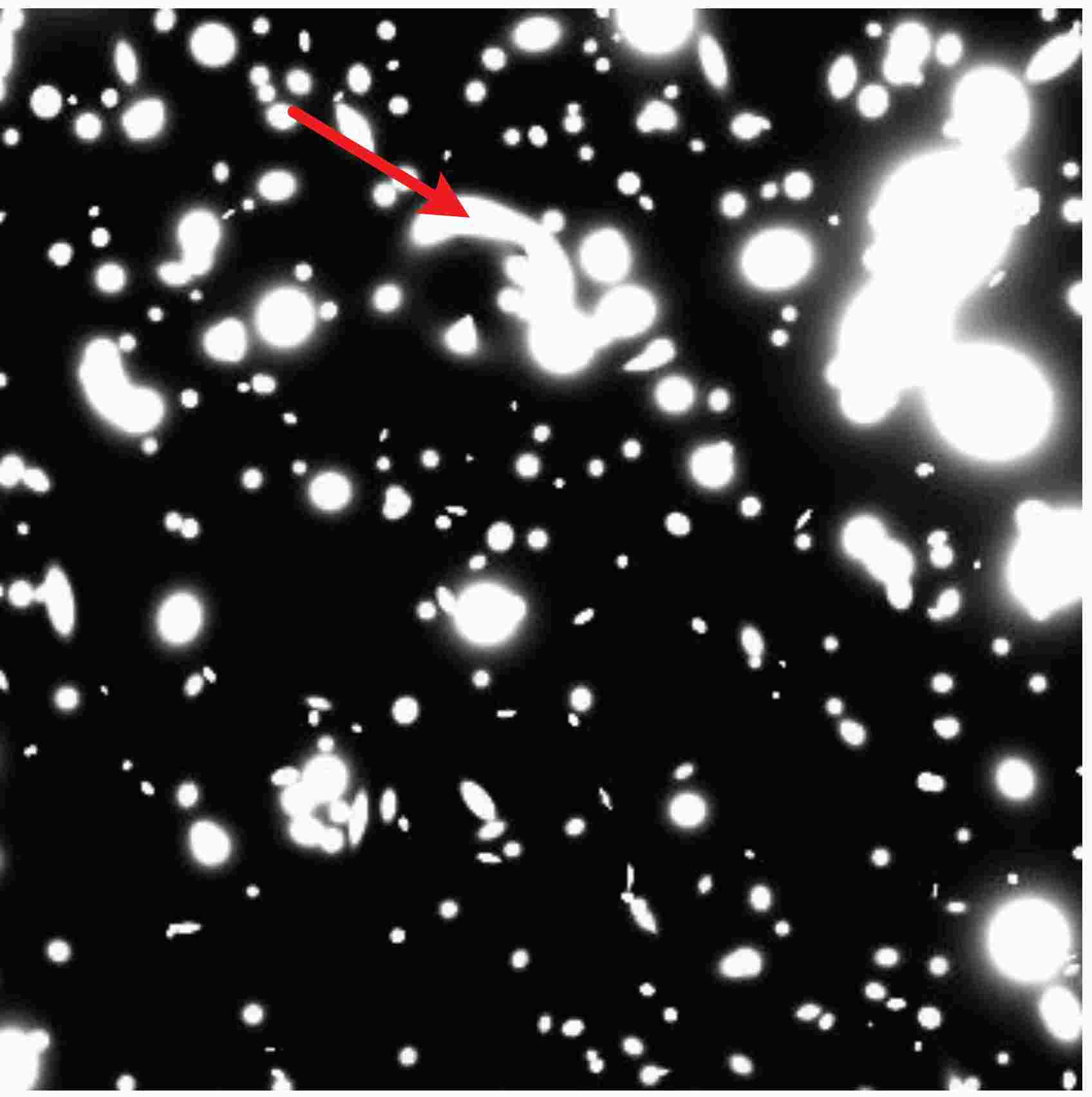}
} 
\caption{This figure shows a simulated image of the same strong gravitational lensing system with and without foreground central galaxies. Figure a and figure b show the same CSGL, but there are obvious foreground central galaxies in figure a and no foreground central galaxies in Figure b. We can easily see morphological characteristics of the CSGL. The arrow in figure a points to the foreground central galaxies. We can see the arc of the strong gravitational lensing system after we remove foreground central galaxies, which is pointed by the arrow in figure b. Therefore, CSGLs with foreground central galaxies interference like figure a pose high requirements and challenges to our detection algorithm.}  
	\label{figure6}  
\end{figure}

\begin{figure}
\centering
\subfigure[]{
\includegraphics[width=0.2\textwidth]{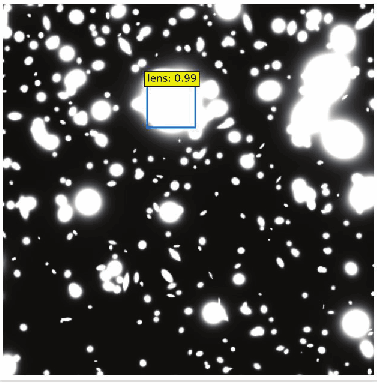}
}
\quad
\subfigure[]{
\includegraphics[width=0.2\textwidth]{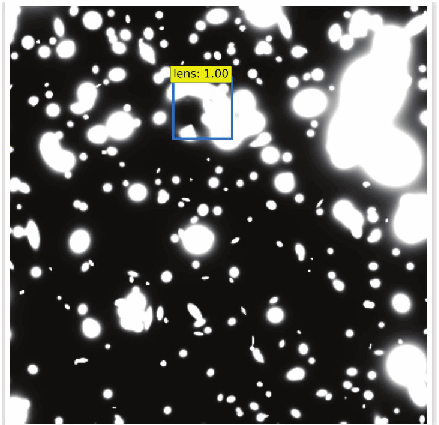}
}
\caption{The results of our algorithm in detection of CSGLs with and without foreground central galaxies. We can see that our algorithm can detect CSGLs regardless of foreground galaxies.}
\label{figure7}
\end{figure}

\begin{figure*}
\centering
\subfigure[]{
\includegraphics[width=0.45\textwidth]{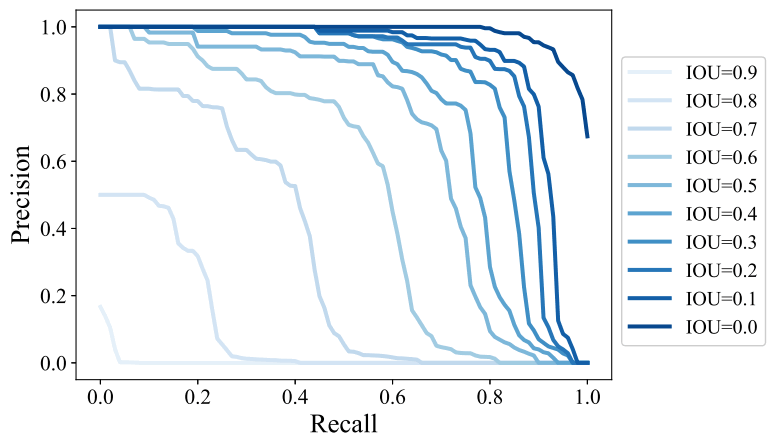}
}
\quad
\subfigure[]{
\includegraphics[width=0.45\textwidth]{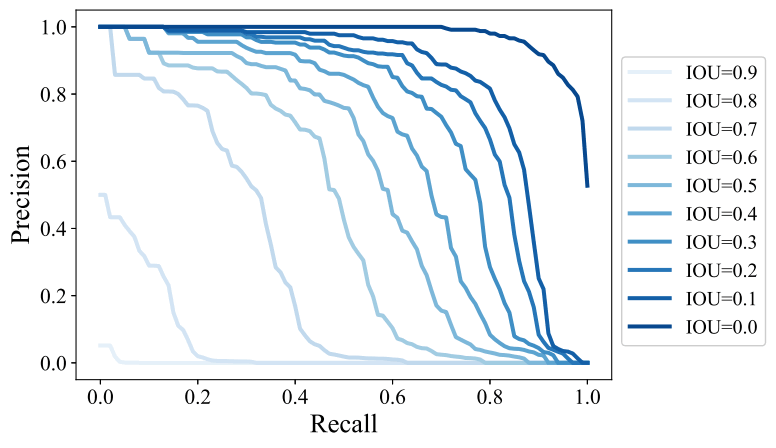}
}\\
\subfigure[]{
\includegraphics[width=0.45\textwidth]{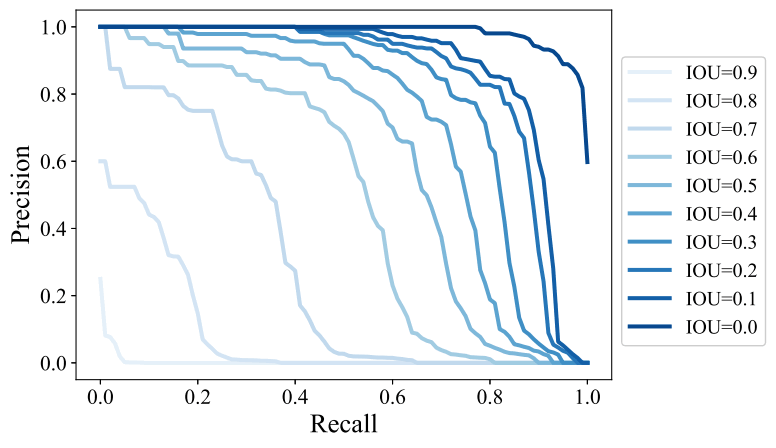}
}
\quad
\subfigure[]{
\includegraphics[width=0.45\textwidth]{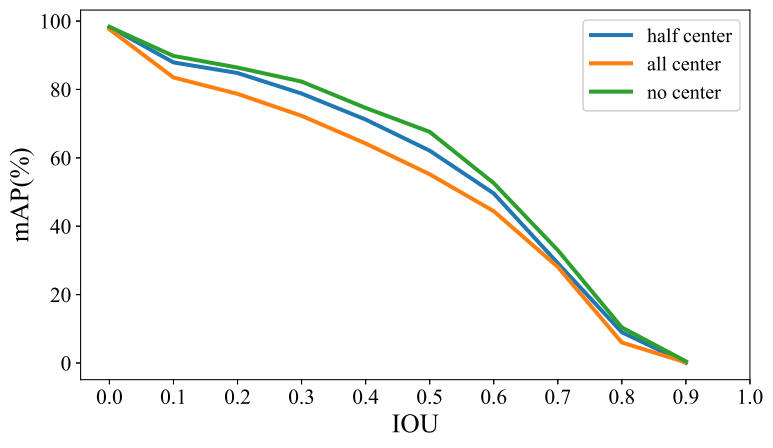}
}
\caption{The performances of our algorithm in detection of CSGLs with and without foreground central galaxies. Figure 9.a, Figure 9.b, and Figure 9.c show the detection ability of our method on data sets with no foreground central galaxies, all of them with foreground central galaxies, and half of them without foreground central galaxies at different IOU thresholds. It can be seen that the model performs best when there is no foreground central galaxies and worst when there are foreground central galaxies, indicating that foreground central galaxies have a certain influence on the performance of our algorithm. In Figure~9.d, the vertical axis represents the area of P-R curve (mAP) for different data sets of the model, and the horizontal axis represents different IOU thresholds. It can be seen that when the IOU threshold is small or large, the difference of the mAP is small, indicating that the foreground central galaxies have little interference to the model at this time. When the IOU threshold is around 0.3, the performance difference of the mAP is the largest. When the algorithm is only used for the classification task of strong gravitational lens, the mAP is all close to $100\%$, which indicates its strong classification ability.}
\label{figure8}
\end{figure*}

Due to the existence of foreground central galaxies, many features of CGSLs, such as arcs and rings are obstructed. Meanwhile, foreground central galaxies will also bring interference to position prediction accuracy of our algorithm. Therefore, we further use simulated images of the same strong gravitational lensing arcs with and without foreground central galaxies to evaluate the performance of our algorithm. A pair of simulated images with and without foreground galaxies are shown in figure~\ref{figure6}. The performances of our algorithm in detection CGSLs with and without foreground central galaxies are shown in figure~\ref{figure7} and figure~\ref{figure8}. In figure~\ref{figure7}, we can find that our algorithm could detect the CGSL, regardless of the foreground galaxies. In figure~\ref{figure8}, we show statistical results of our algorithm in detection CGSLs with or without foreground galaxies. We can find that the difference between detection results is small. Besides, we could find that the performance of our algorithm does not drop significantly when there are BCGs in detection of CGSLs.\\

\section{Detection of CGSLs with the Deformable DETR} \label{sec:4}
For different sky survey projects, the DETR should be trained with simulated data with different observation conditions, such as: observation bands, pixel scales, PSFs and noise levels. However, it would take a long time to train the DETR. Therefore, we propose to use the Deformable DETR for CGSL detection as a light-weighted algorithm. Besides, we further propose to use multi-scale features in the Deformable DETR to increase its accuracy.\\

\subsection{Introduction of the Deformable DETR}
\label{sec:41}
\begin{figure*}
	\centering  
	\includegraphics[width=\textwidth]{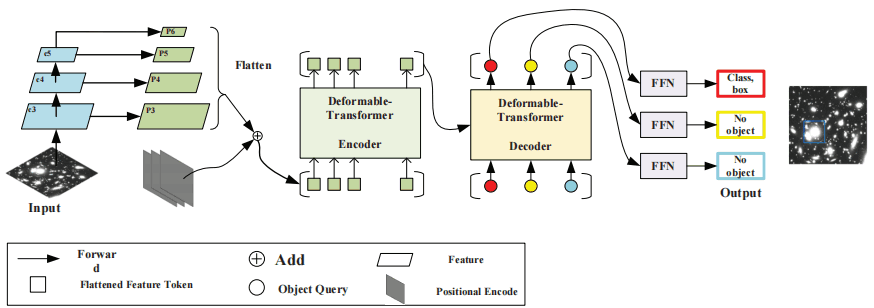}  
	\caption{The schematic diagram of the Deformable DETR. This structural paradigm is the same as that of the DETR with the following differences. First of all, the backbone neural network in the Deformable DETR outputs feature vector maps c3, c4 and c5 from its last three layers. We transform the depth (number of channels) of these three feature vector maps to the same size, thus forming three feature vector maps of p3, p4, and p5 accordingly. Among them, we would convolve feature vector maps of c5 with convolutional layers to further extract feature information and obtain p6 for further processing. We would stack these feature vector maps together to form multi-scale feature, similar to the feature pyramid net (FPN). Secondly, in the encoder-decoder part, the attention mechanism is improved by introducing reference vectors and sampling vectors. Since only the attention map between reference vectors and sampling vectors is calculated, thus greatly reducing the computation complexity and improving the training speed of the model.In this figure, "Flattened Feature Token" represents feature vectors that we input into the transformer. Object queries in different colours represent targets predicted by our model, which is far greater than the actual number of targets in the image. Therefore, in the bipartite matching, most original target instances will be led to "no object", and only a few will be led to corresponding true targets.}  
	\label{figure9}  
\end{figure*}

The DETR algorithm uses feature vectors at all positions to interact with each other for detection, which would require long time and cost a lot of GPU memory. Because CGSLs are sparsely distributed (normally less than one CGSL in an observation image), it is possible to use sparse interactive operations with feature vectors of a few positions for detection. Based on this concept, we propose to use the Deformable DETR for detection. The Deformable DETR \citep{zhu2020deformable} takes advantage of sparse concept from the deformable convolution operation \citep{2017arXiv170306211D} and spatial relation modelling concept from the transformer. The structure of the Deformable DETR is shown in figure~\ref{figure9}.\\

The structure of the Deformable DETR is similar to that of the DETR. But there are two main modifications in the Deformable DETR. Firstly, to keep the performance of the Deformable DETR stable, we extract several instead of all feature vectors from last few layers in Resnet-50 as the multi-scale feature, which is similar to feature pyramid network (FPN) used in celestial objects detection for wide field small aperture telescopes \citep{2020AJ....159..212J}. In this way, the model can obtain more features of input images at different scales and levels, which could help the model to obtain richer feature information.Secondly, we introduce the attention mechanism operation with adjacent feature vectors in the Deformable DETR. A feature vector will interact only with adjacent feature vectors, which could reduce complexity. This feature vector is called the reference vector, while these adjacent feature vectors sampled by the Deformable DETR are called sampling vectors. Meanwhile, since the size of CGSLs is large and sparsely distributed, sparsely distributed sampling vectors would not seriously affect the performance. Sampling features would be sent to attention layers of the decoder. After data propagation through these layers to attention layers of the decoder, multi-scale attention operation would be carried out to give final results.\\

\subsection{Performance Comparison between the DETR and the Deformable DETR}
\label{sec:42}
\begin{figure*}
\centering
\includegraphics[width=0.45\textwidth]{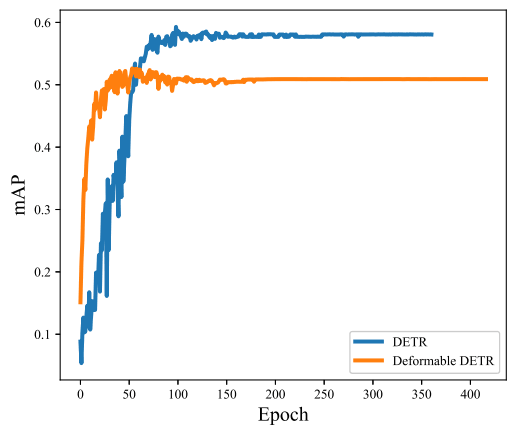}
\caption{The learning curve of the DETR and that of the Deformable DETR. As we can find in this figure, the Deformable DETR will converge after around 100 epochs and the DETR will require 200 epochs to converge. However, the DETR has a higher mAP after training.}
\label{figure10}
\end{figure*}

\begin{figure*}
\centering
\includegraphics[width=0.45\textwidth]{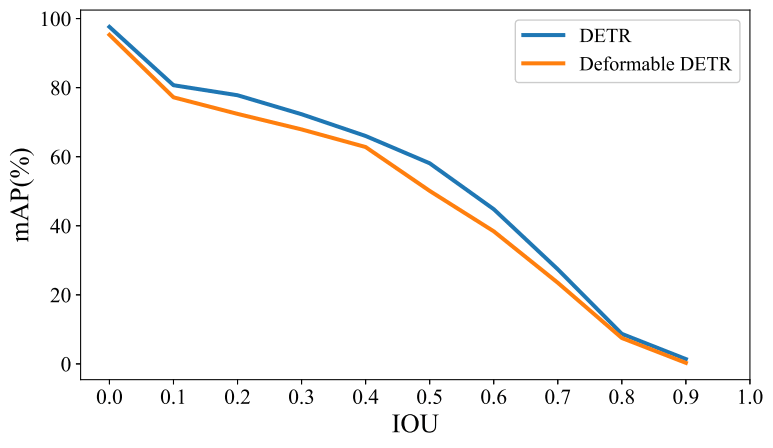}
\caption{The mAP of the DETR and that of the Deformable DETR with different IOU criterion. As shown in this figure, we can find that performance of the DETR is better than that of the Deformable DETR. }
\label{figure11}
\end{figure*}

\begin{figure*}
\centering
\subfigure[]{
\includegraphics[width=0.2\textwidth]{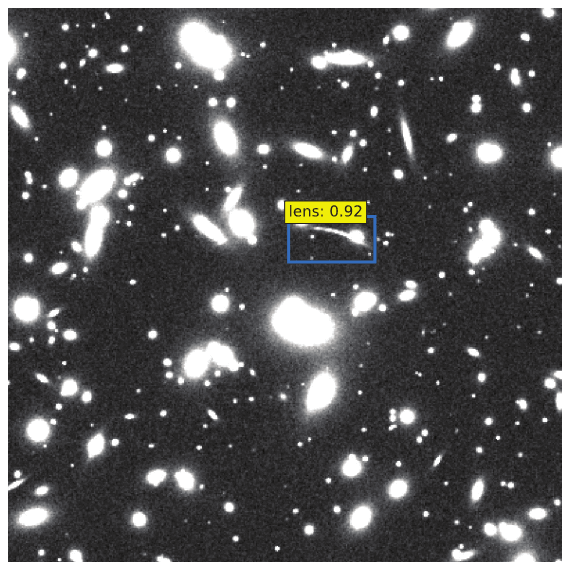}
}
\quad
\subfigure[]{
\includegraphics[width=0.2\textwidth]{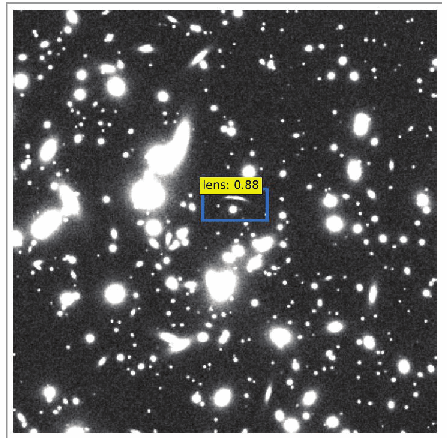}
}
\subfigure[]{
\includegraphics[width=0.2\textwidth]{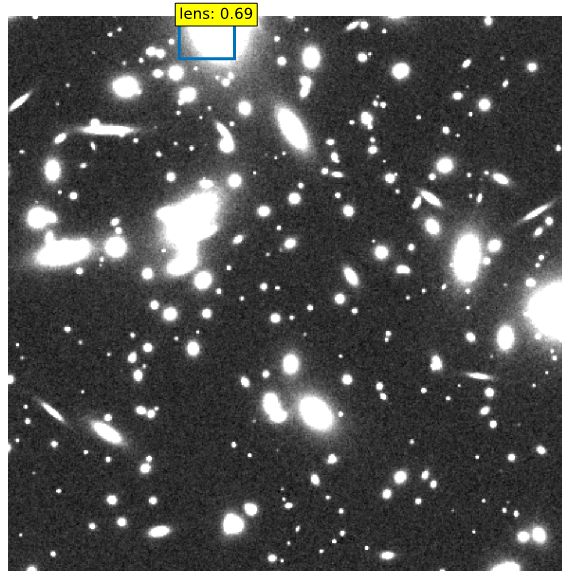}
}
\quad
\subfigure[]{
\includegraphics[width=0.2\textwidth]{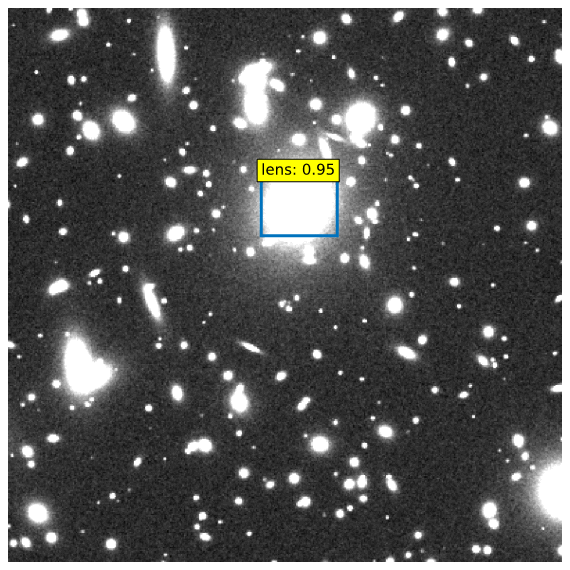}
}\\
\subfigure[]{
\includegraphics[width=0.2\textwidth]{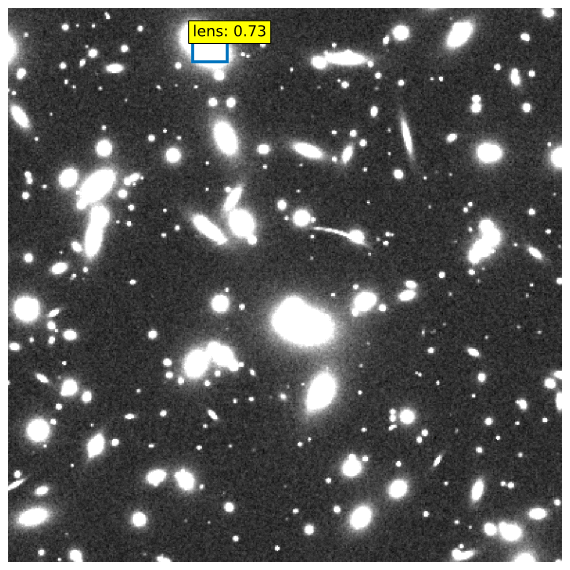}
}
\quad
\subfigure[]{
\includegraphics[width=0.2\textwidth]{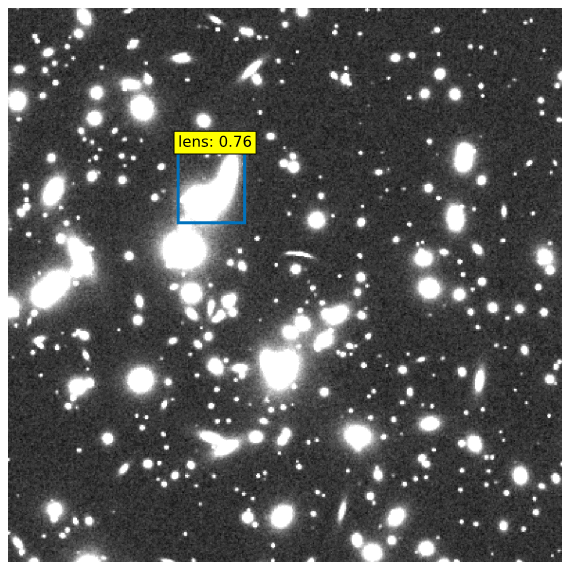}
}
\subfigure[]{
\includegraphics[width=0.2\textwidth]{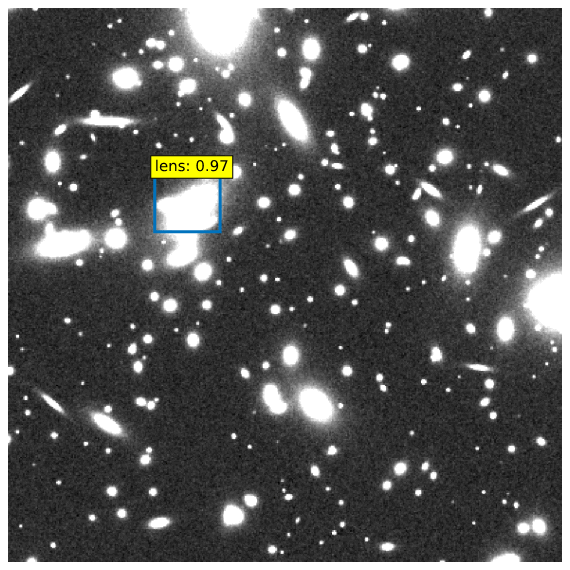}
}
\quad
\subfigure[]{
\includegraphics[width=0.2\textwidth]{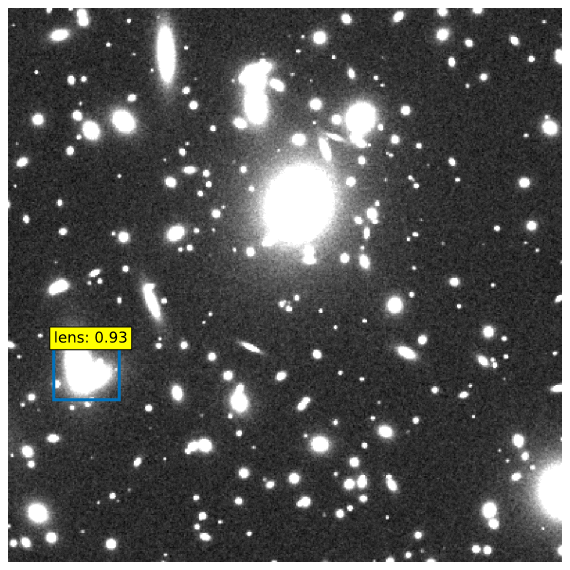}
}

\caption{This figure shows detection results of the two different methods in four figures. Figure \ref{figure12}.a \ref{figure12}.b \ref{figure12}.c \ref{figure12}.d show the detection results of the DETR and figure \ref{figure12}.e \ref{figure12}.f \ref{figure12}.g \ref{figure12}.h show the detection results of the Deformable DETR. We can find that the DETR gives correct results for figure \ref{figure12}.a and \ref{figure12}.b, while the Deformable DETR could give correct results for figure \ref{figure12}.g and \ref{figure12}.h. Meanwhile, for figure \ref{figure12}.g and \ref{figure12}.h, we find that the arc is obscured by the foreground galaxy, so it is difficult for eyes to see it, but our algorithm can effectively identify and locate it.}
\label{figure12}
\end{figure*}

We train the Deformable DETR with the same dataset as we used in Section \ref{sec:32}. The optimizer and the loss function are the same as the DETR. The model training process is shown in figure~\ref{figure10}. With about 100 epochs, the Deformable DETR would converge. However, the DETR needs about 200 epochs to converge. Therefore the convergence speed of the Deformable DETR is greatly improved. But there is a slightly decrease in detection performance. Figure~\ref{figure5} and Figure~\ref{figure11} shows the detection performance of these two algorithms under different IOU thresholds. It can be seen that the overall performance of Deformable DETR is lower than that of DETR, because the introduction of the deformable attention mechanism effectively reduces the amount of computation, but also loses the information of non-reference vectors and non-sampling vectors. Still, we can see that the performance degradation is small, reaching a maximum of no more than 10 percent at the IOU threshold of 0.5. We select several CGSLs that are detected by the DETR and missed by the Deformable DETR and also some CGSLs that are detected by the Deformable DETR and missed by the DETR in Figure~\ref{figure12}. From these figures, we can find that the DETR could detect arc structures and might be affected by foreground galaxies. The Deformable DETR would occasionally detect CGSLs with foreground galaxies when reference points are selected in some priority positions, which would better extract features of CGSLs for detection.\\

\subsection{Increasing the Performance in Detection of CGSLs with Ensemble Learning for Real Applications}
\label{sec:43}
Since we could find that the DETR and the Deformable DETR have different designs and their performance is different for the same dataset. Two machine learning algorithms with different performance could be merged together with resemble learning to further improve their performance. With this concept, we use equation~\ref{equation6} to calculate final results with detection results of the DETR and that of the Deformable DETR,\\
\begin{equation}
		out = out_{max}(Score_{DETR},Score_{DeformableDETR}),\\
		\label{equation6}
\end{equation}
where $Score_{DETR}$ and $Score_{DeformableDETR}$ are scores of the DETR and that of the Deformable DETR. With equation~\ref{equation6}, we would output candidates with high confidence of either of these two algorithms. For comparison, the DETR is used as baseline for comparison as shown in table~\ref{table1}. The results of the ensemble learning algorithm are shown in table~\ref{table2}. We can find that with IOU of 0.1 and Score of 0.7, the DETR has Recall rate of $66.0 \%$ and Precision rate of $78.9 \%$, while the ensemble learning algorithm has Recall rate of $88.0 \%$ and Precision rate of $70.2 \%$. More CGSLs could be detected with ensemble learning with only small drop of precision rate. For rare target detection, such as CGSLs, the ensemble learning would be a better choice.\\

	\begin{table*}
	\centering
	\caption{The recall rate and the precision rate (R/P) of the DETR with different IOUs and Scores. }
	\begin{tabular}{ccccccccc} 
	\toprule
	\diagbox{IOU}{R/P}{Score} & 0.9       & 0.85      & 0.8       & 0.75      & 0.7       & 0.65      & 0.6       & 0.55       \\ 
	\hline
	\multicolumn{1}{c}{0.9}   & 0.0/0.0   & 0.0/0.0   & 0.0/0.0   & 0.0/0.0   & 0.0/0.0   & 0.0/0.0   & 0.0/0.0   & 0.0/0.0    \\
	\multicolumn{1}{c}{0.8}   & 4.9/5.9   & 5.8/5.0   & 7.5/4.7   & 8.2/4.5   & 9.4/4.2   & 10.5/4.0  & 12.1/3.8  & 13.6/3.7   \\
	\multicolumn{1}{c}{0.7}   & 13.7/18.3 & 16.3/15.8 & 20.9/15.4 & 30.0/15.0 & 26.1/14.4 & 28.7/13.7 & 32.0/13.1 & 35.0/12.6  \\
	\multicolumn{1}{c}{0.6}   & 23.5/35.5 & 27.4/30.7 & 33.9/30.0 & 37.1/29.6 & 41.7/29.1 & 45.5/28.4 & 49.4/27.1 & 52.8/26.2  \\
	\multicolumn{1}{c}{0.5}   & 32.6/55.9 & 39.8/53.8 & 46.8/51.4 & 49.8/49.8 & 54.9/49.5 & 58.4/47.8 & 62.5/46.2 & 65.8/44.9  \\
	\multicolumn{1}{c}{0.4}   & 37.5/69.4 & 45.1/67.0 & 52.9/65.6 & 56.1/64.0 & 60.8/63.2 & 64.1/60.9 & 68.2/60.0 & 71.3/58.1  \\
	\multicolumn{1}{c}{0.3}   & 40.1/77.4 & 47.8/74.7 & 55.8/73.9 & 59.0/72.3 & 63.8/71.2 & 67.1/69.6 & 71.1/68.2 & 74.1/66.8  \\
	\multicolumn{1}{c}{0.2}   & 41.2/82.3 & 49.3/79.2 & 57.2/78.3 & 60.4/76.4 & 65.1/75.8 & 68.5/74.2 & 72.6/73.2 & 75.5/72.0  \\
	\multicolumn{1}{c}{0.1}   & 42.2/84.4 & 49.9/81.0 & 58.0/80.6 & 61.3/79.4 & 66.0/78.9 & 69.6/77.9 & 73.6/77.1 & 76.4/75.7  \\
	\multicolumn{1}{c}{0.0}   & 42.8/86.6  & 50.7/83.7  & 58.8/83.4  & 62.1/82.4  & 66.8/81.8  & 70.5/81.6  & 74.4/80.6  & 77.2/79.4   \\
	\bottomrule
	\end{tabular}
	\label{table1}
	\end{table*}

	\begin{table*}
	\centering
	\caption{The recall rate and the precision rate (R/P) of the Ensemble model with different IOUs and Scores.}
	\begin{tabular}{ccccccccc} 
	\toprule
	\diagbox{IOU}{R/P}{Score} & 0.9       & 0.85      & 0.8       & 0.75      & 0.7       & 0.65      & 0.6       & 0.55       \\ 
	\hline
	\multicolumn{1}{c}{0.9}   & 0.0/0.0   & 0.7/0.3   & 1.0/0.3   & 1.5/0.3   & 2.8/0.3   & 5.0/0.3   & 9.0/0.3   & 14.3/0.3    \\
	\multicolumn{1}{c}{0.8}   & 6.0/5.6   & 8.8/4.9   & 12.5/4.6   & 17.5/4.2   & 28.6/3.8  & 42.4/3.7  & 58.3/3.6   & 70/3.5   \\
	\multicolumn{1}{c}{0.7}   & 16.8/17.8 & 23.7/15.8 & 31.0/14.5 & 40.5/13.4 & 56.8/12.6 & 70.8/12.0 & 82.1/11.7   & 88.5/11.6  \\
	\multicolumn{1}{c}{0.6}   & 28.0/34.3 & 36.6/29.3 & 45.6/27.1 & 56.9/26.0 & 72.0/24.6 & 82.9/24.1 & 90.2/23.5 & 93.9/23.3  \\
	\multicolumn{1}{c}{0.5}   & 36.3/50.2 & 48.1/47.0 & 57.2/43.2 & 67.5/40.9 & 80.2/38.8 & 88.3/37.7 & 93.5/36.8 & 96.0/36.5  \\
	\multicolumn{1}{c}{0.4}   & 42.3/64.8 & 54.5/60.9 & 63.6/56.4 & 73.3/54.0 & 84.2/51.1 & 90.9/49.5 & 95.0/48.6 & 97.0/48.4  \\
	\multicolumn{1}{c}{0.3}   & 44.9/71.8 & 57.1/67.7 & 66.6/64.4 & 76.0/62.4 & 86.1/59.0 & 92.1/57.9 & 95.7/57.0 & 97.4/56.7  \\
	\multicolumn{1}{c}{0.2}   & 47.2/78.9 & 59.2/73.7 & 68.7/71.0 & 77.8/69.0 & 87.2/65.3 & 92.8/64.4 & 96.1/63.4 & 97.7/63.3  \\
	\multicolumn{1}{c}{0.1}   & 48.6/83.6 & 60.5/77.8 & 69.9/75.2 & 78.8/73.1 & 88.0/70.2 & 93.4/69.9 & 96.4/69.1 & 97.8/68.6  \\
	\multicolumn{1}{c}{0.0}   & 49.2/85.4  & 61.3/80.5  & 70.8/78.5  & 79.4/76.1  & 88.6/74.1  & 93.7/73.8  & 96.6/72.9  & 97.9/72.4   \\
	\bottomrule
	\end{tabular}
	\label{table2}
	\end{table*}

\section{Interpretation of the Mechanism of the attention based CGSL detection algorithm}
\label{sec:5}
Both the DETR and the Deformable DETR use the attention mechanism to detect targets. The attention mechanism is quite similar to the attention of human-beings in observing a target. In this part, we would draw attention maps in different attention layers to show how the DETR and the Deformable DETR work and help us to better evaluate their performance. There are two types of attention maps from two different attention layers that we red are going to visualize: the self-attention map in the last self-attention layer of the encoder part and the cross-attention map in the last cross-attention layer of the decoder part. We will show both of these two attention maps for two scenarios: a CGSL which is obstructed by the foreground galaxies and a CGSL which is not obstructed by the foreground galaxies.\\

\begin{figure*}
\centering
\subfigure[]{
\includegraphics[width=0.87\textwidth]{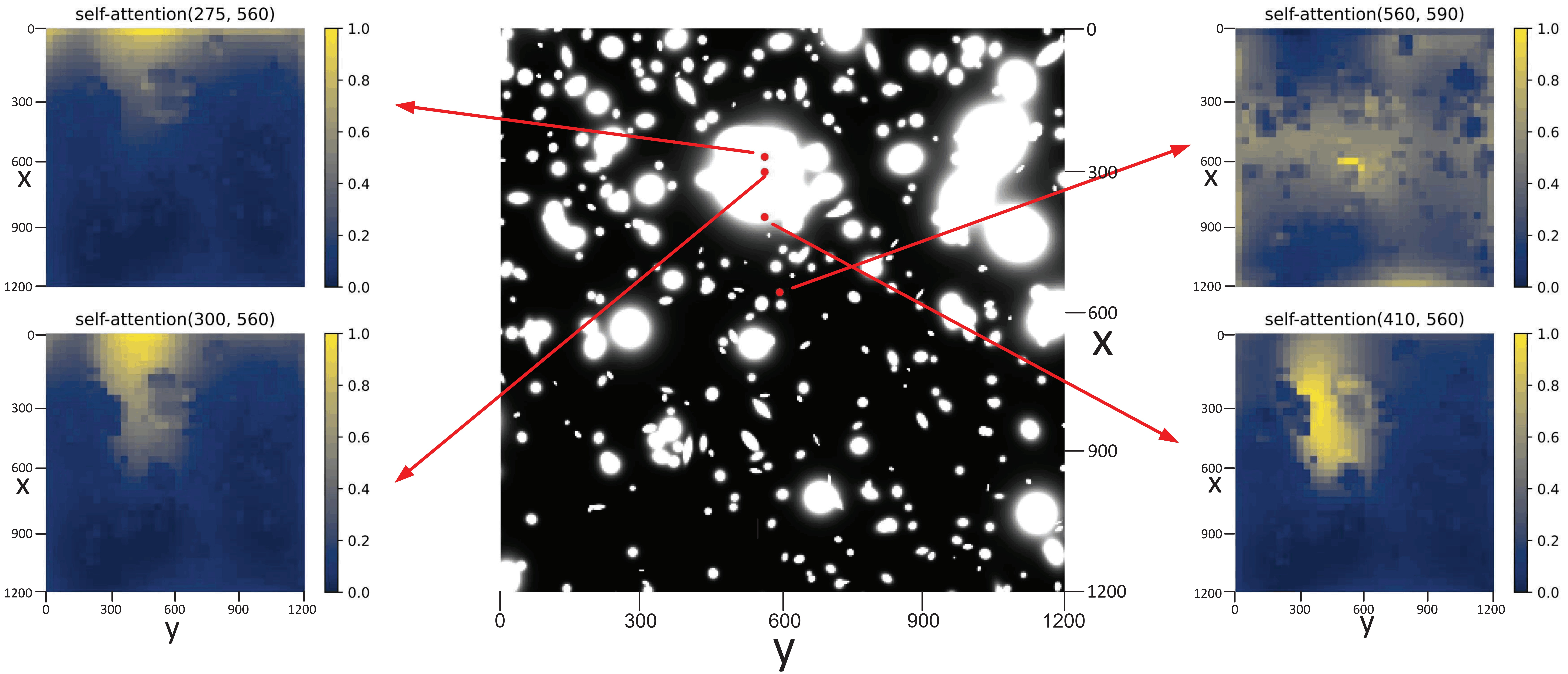}
}
\quad
\subfigure[]{
\includegraphics[width=0.87\textwidth]{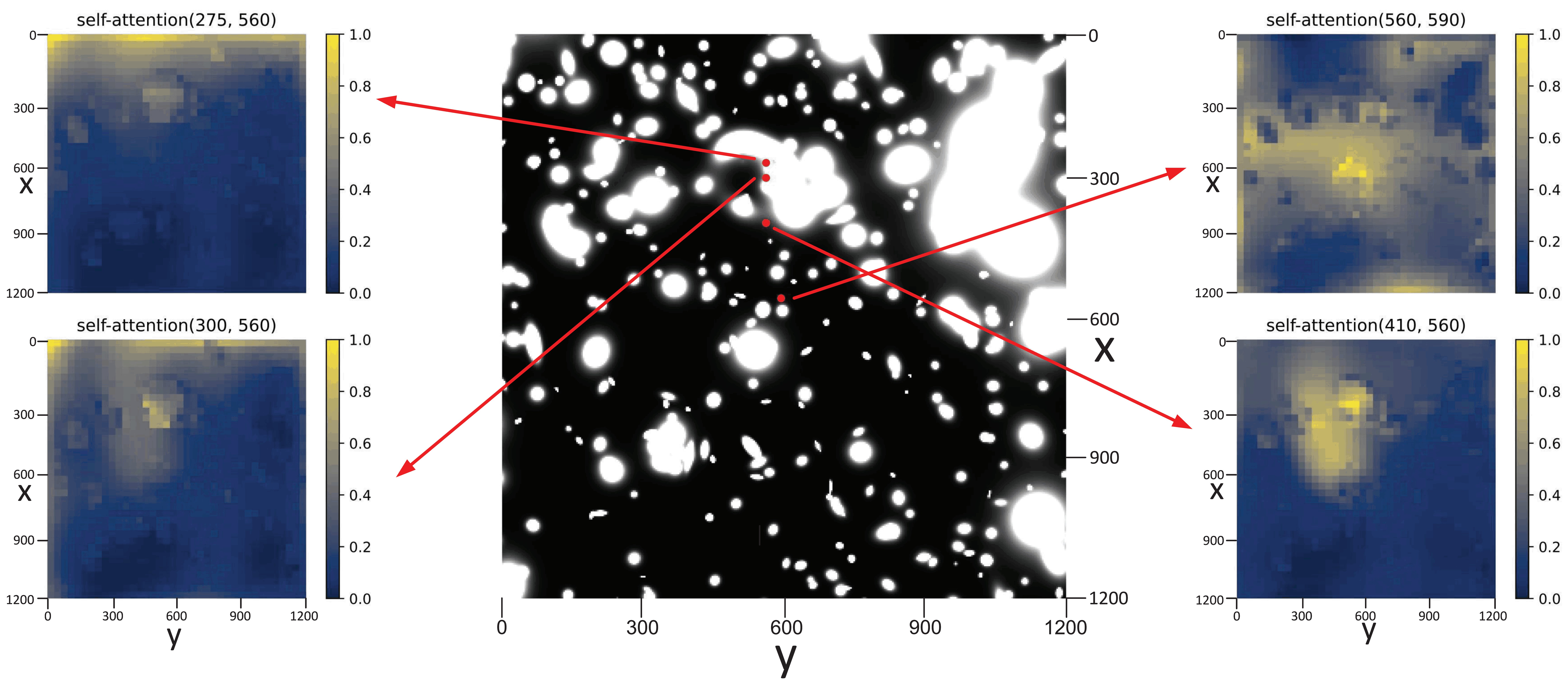}
}
\caption{The self-attention map of four pixels in the last layer of the encoder. Figure 14.a and figure14.b are CGSLs with and without foreground central galaxies respectively. We can find that the self-attention maps of them are similar, which indicates that our model can still identify CGSL features accurately and has good robustness even in the presence of foreground central galaxies. It also shows that the encoder is paying attention to the main structure of the CGSL, regardless of BCGs.}
\label{figure13}
\end{figure*}

We will firstly show self-attention maps in the last self-attention layer of the encoder part. When we input an image into the model, the weight of the relation between feature vectors of different pixels in the image will be calculated in the last self-attention layer of the encoder as the attention map. If we select feature vectors from one pixel and calculate relations between other pixels, we would get an attention map with reference to that pixel. Therefore, we could select several pixels in the input image as sampling pixels and get their attention maps in the encoder's last self-attention layer. Then we can visualize the relationship between sampling pixels and other pixels in the input image to indicate us the attention of the detection algorithm. As shown in Figure~\ref{figure13}, we select four pixels in the image to show their attention maps. We use the right-handed Cartesian coordinate system to define coordinates of these four pixels and use red dots to indicate their positions in the figure. Figure~\ref{figure13}.a and Figure~\ref{figure13}.b are CGSLs with and without foreground central galaxies respectively. \\

In figure~\ref{figure13}.a, we can see that when the sampling pixel (560, 590) is not in the CGSL, the attention distribution is uniform and random, indicating that there is no clear correlation or significant difference between the current sampling pixel and other pixels. Therefore, the sampling pixel does not participate in the detection of CGSLs and this pixel does not belong to the image of the CGSL. Sampling pixels (275,560), (300, 560), (410, 560) are inside the CGSL. We can see that as sampling pixels approach to the CGSL, the distribution of the attention map gradually concentrates to a circle around the CGSL. It indicates us that there exists higher attention weight between the sampling pixel and other pixels within the CGSL. This strong relationship shows that these pixels are more closely related to each other, which could be used to characterize the CGSL system. The self-attention map within the CGSL is not exactly the same between different sampling pixels. It can be seen that the concentration of attention distribution of  sampling pixels at the edge of the CGSL system is relatively poor, and there is a certain correlation with other irrelevant pixels around, which reflects that pixels at the edge of the CGSL image have weaker effects in obtaining final detection results. The concentration of attention distribution of sampling pixels in the centre of the CGSL is very high in self-attention map, indicating that these pixels are important in detection of CGSLs. It should be noted that for labels in the training dataset, only the boundary information of the target is given instead of detailed pixel-level information. However, in self-attention maps, we can find that our algorithm could obtain pixel-level information of CGSLs, although in a relatively rough way. It still shows that thanks to the attention mechanism, our algorithm can learn and capture some important non-human and non-prior features of the target by itself, and achieve simple image segmentation results. In figure~\ref{figure13}.b, we can draw a similar conclusion. Whether or not the CGSL is obstructed by foreground BCGs, we can output an effective self-attention map, which indicates the effectiveness of our method. We can also find that the self-attention maps of figure~\ref{figure13}.a and figure~\ref{figure13}.b are similar, which indicates that our model can still identify CGSL features accurately and robustly even in the presence of foreground central galaxies. Overall, figure~\ref{figure13}.a and figure~\ref{figure13}.b maintain high similarity. \\

\begin{figure}
\centering
\subfigure[]{
\includegraphics[width=0.85\textwidth]{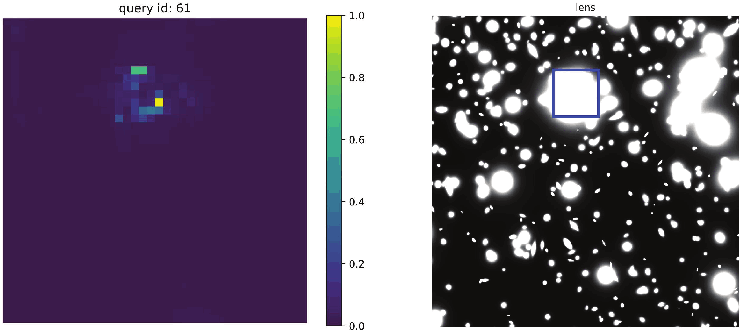}
}
\quad
\subfigure[]{
\includegraphics[width=0.85\textwidth]{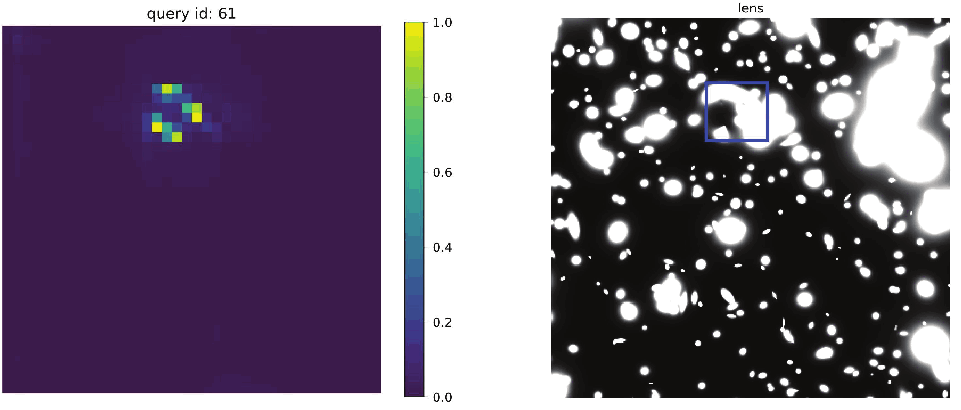}
}
\caption{The cross-attention map of the last layer in the decoder. Input images are the same as those in figure~\ref{figure13}. In this figure, right subfigures are prediction results and left subfigures are cross-attention maps of object queries matching predicted results in the decorder, whose legend represents the serial number of the object query. We can see that the decoder pays more attention to boundaries of CGSLs and gives similar results regardless of the interference of foreground central galaxies.}
\label{figure14}
\end{figure}

Then we visualize the cross-attention map in the last cross-attention layer of the decoder part. We directly use cross-attention map of predicted results to visualize attentions. As we have mentioned earlier, each object query will predict a target, and only a few of them are real targets.  Here, we will show queries that match real targets. Then we could obtain outputs of these queries in the last decorder layer and visualize it to obtain the cross-attention map of predicted targets. As shown in figure~\ref{figure14}, compared with the self-attention encoder which focuses on the main part of the object, the decoder based on the cross-attention mechanism pays more attention to boundaries of CGSLs. Figure~\ref{figure14}.a and figure~\ref{figure14}.b are CGSLs with foreground central galaxies and without foreground central galaxies respectively. We find that the cross-attention map pays more attention to CGSL boundaries than other parts. Meanwhile, for the same CGSL system, regardless of whether there is a foreground central galaxy or not, the cross-attention map is very similar, which indicates that our method is robust in detection of CGSLs.\\

With visualization methods mentioned above, we could find that attention based detection algorithms firstly pay attention to the main body of CGSLs with self-attention layer in the encoder. Then the detection algorithms pay attention to edge of CGSLs with cross-attention layer in the decoder part. This is similar to the attention mechanism by which people observe things. However, attention based detection algorithms could directly use CGSL images of  multiple bands for detection. Although the image of the CGSL is strongly obstructed by the foreground galaxies, these detection algorithms could still detect the CGSLs. Thanks to this property, attention based detection algorithms could achieve better performance in detection of CGSLs than that could be obtained by humans and would lead to new discoveries of CGSLs.\\

\section{Performance Test with Simulated and Real Observation Data}
\label{sec:6}
\subsection{Performance Test with Simulated Data}
\label{sec:61}
\begin{figure*}
\centering
\subfigure[]{
\includegraphics[width=0.85\textwidth]{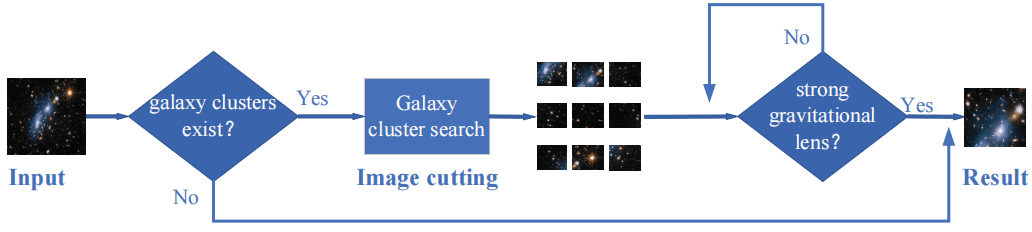}
}
\quad
\subfigure[]{
\includegraphics[width=0.65\textwidth]{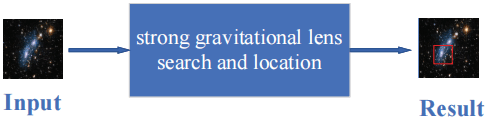}
}
\quad
\subfigure[]{
\includegraphics[width=0.85\textwidth]{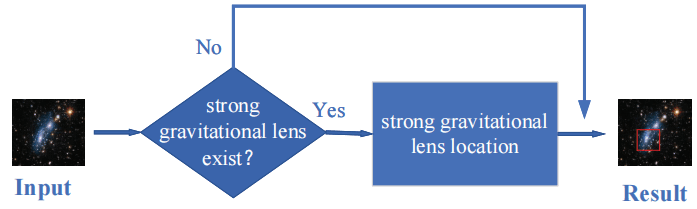}
}
	\caption{This figure shows the schematic diagram of three different approaches. Figure a shows the classic method, which firstly detects images of galaxy clusters. Then these stamp images would be cut from original observation images and classified either to CGSLs or other targets, as discussed in \citet{2017MNRAS.471..167J,2017MNRAS.472.1129P,2017MNRAS.465.4325O,2017A&A...597A.135B,2017MNRAS.471.3378H,2019ApJ...877...58A,2018MNRAS.473.3895L}. Figure b shows the diagram to use our algorithm in a one step way, which directly detect CGSLs from observational images. Figure c shows the two step strategy. We firstly obtain candidates with low IOU and then detect CGSLs from these images.}
	\label{figure21}  
\end{figure*}

In this subsection, we will consider to use our method in real applications. Since CGSLs are rare in real conditions and detection results would be analyzed by scientists for further study, it would be more appropriate to design a framework with high precision, moderate completeness and low false positive rate. Considering the recall rate is more important than position accuracy in practical applications, we do not need to use very high position accuracy conditions all the time. Firstly, we use a low IOU to classify full frame images and obtain candidate images that may contain CGSLs. In this way, the position accuracy is low, but the recall rate and precision rate are high. Then we use a high IOU to detect the candidate images again, so as to accurately find the CGSLs in these candidate images, which ensures the position accuracy. The flowchart of the classic method, our previous method and the two step strategy are shown in figure~\ref{figure21}. We would firstly use the ensemble learning framework discussed above to detect CGSLs from observational images and we would further use the following strategy to obtain final detection results. Detail stpng are shown below:\\
1. We set the IOU as 0.0 and obtain images that may contain CGSLs as candidates.\\
2. We set the IOU as 0.7 and detect CGSLs from these candidate images with either the deformable DETR or the DETR, whose detection result is adapted in the previous step.\\

\begin{figure*}
\centering
\includegraphics[width=0.5\textwidth]{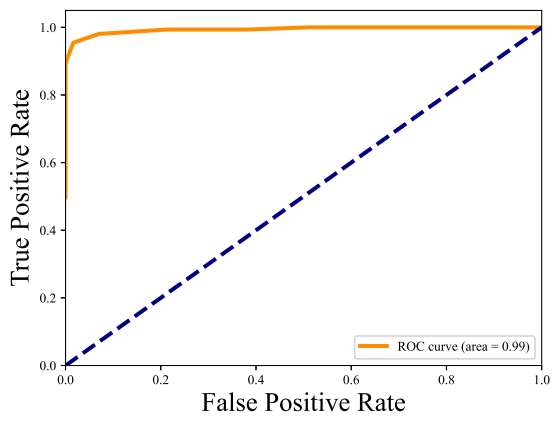}
\caption {The ROC curve of our algorithm, which has an area of 0.99.}
\label{figure23}  
\end{figure*}

To test the performance of the two step approach in real applications, we have generated simulated data with 16,000 images and $1 \%$ of them contain CGSLs. It would cost about 180 minutes to process all these images. The test results show that our method can achieve $99.63 \%$ accuracy, when the recall rate is $90.32\%$ and the precision is $87.53 \%$. Besides the false positive rate is only $0.23 \%$. We have further plotted ROC curve as shown in figure~\ref{figure23}. The low false positive rate ensures that future applications of large-scale sky surveys will not produce a lot of false positives, which means there won't be too much reliance on human inspection. Although the number of candidates to be checked by scientists has been greatly reduced, there are still lots of candidates need to be checked by human investigation. However, as we have discussed in Section \ref{sec:2}, our algorithm could detect CGSLs that could not be directly seen by human. Therefore, we need to integrate the attention map with observation images to generate augmented images and use citizen science platforms to investigate details of these images \citep{lintott2008galaxy, smith2011galaxy, brink2013using, marshall2015ideas}. We would further investigate this problem in our future works.

\subsection{Performance Test with Real Observation Data}
\label{sec:62}
\begin{figure*}
\centering
\subfigure[]{
\includegraphics[width=0.26\textwidth]{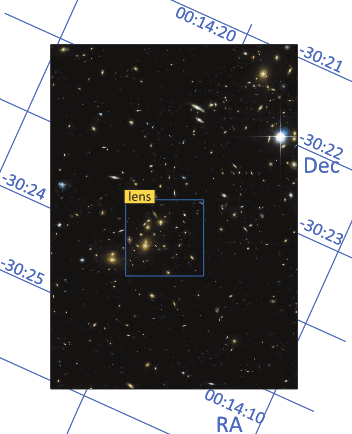}
}
\quad
\subfigure[]{
\includegraphics[width=0.3\textwidth]{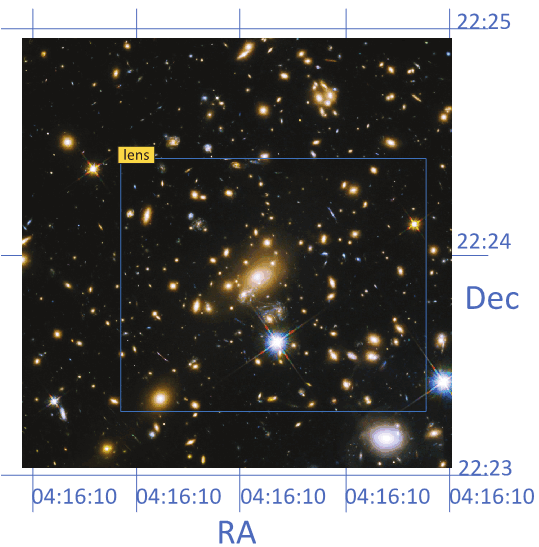}
}
\quad
\subfigure[]{
\includegraphics[width=0.32\textwidth]{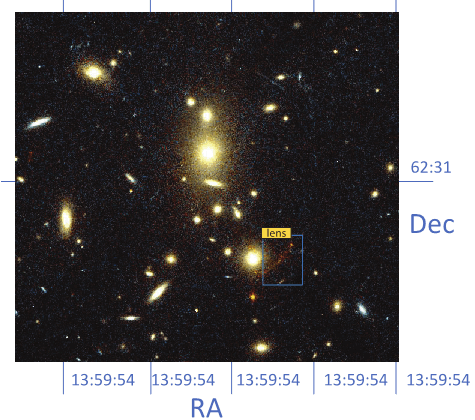}
}
\caption{Detection results of our algorithm for images contain Abell 2744  ($3.05\times 4.38$ arcmin with $3662\times 5253$ pixels), MACS J1149.5+2223 ($1.90\times 1.90$ arcmin with $3800\times 3800$ pixels) and ClG 1358+62 ($1.06\times 0.96$ arcmin with $640\times 583$ pixels).  These results show that our algorithm could directly detect gravitationally arcs from real observation data, when our algorithm is only trained with ideal simulated data without noise. Generally speaking, a model trained only with simulated data usually needs some methods to adapt to real observation data. However, our algorithm can detect strong gravitational lensing on real data without further training, which shows it is effective to train an effective neural network with simulated data, if the simulation reflects real physical process. Original image of Abell 2744 by NASA, ESA and D. Coe (STScI)/J. Merten (Heidelberg/Bologna). Original image of MACS J1149.5+2223 by NASA, ESA, S. Rodney (John Hopkins University, USA) and the FrontierSN team; T. Treu (University of California Los Angeles, USA), P. Kelly (University of California Berkeley, USA) and the GLASS team; J. Lotz (STScI) and the Frontier Fields team; M. Postman (STScI) and the CLASH team; and Z. Levay (STScI). Original image of CIG 1358+62 by Marijn Franx (University of Groningen, The Netherlands), Garth Illingworth (University of California, Santa Cruz) and NASA/ESA.}
\label{figure15}
\end{figure*}

We use real observation data obtained from the Frontier Fields survey project and the RELICS survey project obtained by the Hubble Space Telescope and the early release deep field image from the James Webb Space Telescope to test the performance of our algorithm. Firstly, we select several images obtained by the frontier Fields survey project \citep{2014AAS...22325401L, 2014AAS...22325402K, 2017ALotzpJ}  to test the performance of our algorithm. These images are downloaded from the website (\url{https://esahubble.org/}) as Fullsize Original TIF files. We directly use the ensemble learning algorithm to connect the DETR and the Deformable DETR trained with simulated data to detect CGSLs in these images. Since these images does not have any bounding boxes, we check these images by eyes and use the dark matter distribution released in the same website as reference to locate CGSLs. We have tested our algorithm with all these images, which include images of Abell2744, MACS J0416.1-2403, MACS J0717.5+3745, MACS J1149.5+2223, Abell S1063 and Abell 370. Generally speaking, our algorithm could detect almost all gravitationally arcs from these images, as shown in figure \ref{figure15}.\\

\begin{figure*}
\centering
\subfigure[]{
\includegraphics[width=0.32\textwidth]{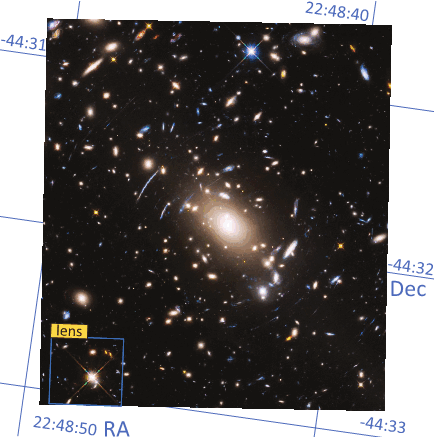}
}
\quad
\subfigure[]{
\includegraphics[width=0.3\textwidth]{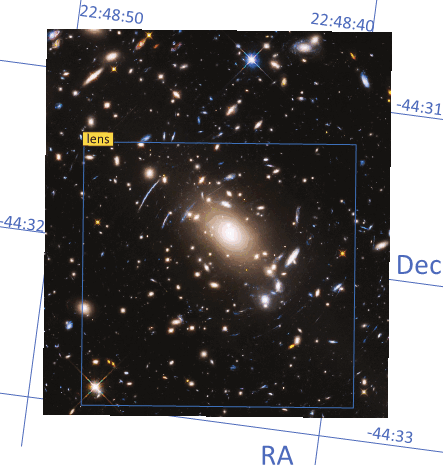}
}
\caption{Detection results for images contain Abell S1063($2.07\times 2.32$ arcmin) in images of different size, $2243\times 2511$ pixels in the left figure and $1280\times 1433$ pixels in the right figure. A star with a bright arc nearby is wrongly classified as a CGSL, because it has relatively large size and contains a central bright source and arcs nearby. While our algorithm could detect the CGSL from an image with smaller size, when the star and the arc has smaller size. Original image of Abell S1063 by NASA, ESA, and J. Lotz (STScI).}
\label{figure16}
\end{figure*}

Difference between training data and real data would introduce problems to machine learning algorithms, when they are deployed. Although our algorithm has shown relative good generalization performance and robustness in figure \ref{figure15}, its performance is still affected. For CGSL detection task, images that have similar arc structures might introduce errors to detection results. Since the training data does not contain any diffraction effects brought by the optics system, our detection algorithm has some risks to detect `arc' structures brought by the diffraction effect as shown in figure \ref{figure16}.a. We can find that our detection algorithm locates a star with an arc structure nearby. However, if we use our algorithm to detect CGSLs in the same image with lower resolution (smaller number of pixels here), our algorithm could detect CGSLs as shown in figure \ref{figure16}.b. The problem of wrong detections brought by diffraction effects could be solved if we add more realistic effects to simulated images as training data, or we use real observation data to train the neural network with transfer learning strategy. \\

\begin{figure}
\centering
\includegraphics[width=0.4\textwidth]{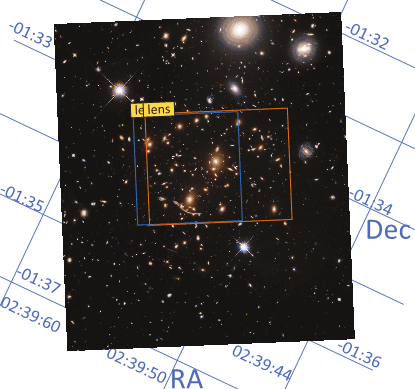}
\caption{Detection results with top 2 scores for an image which contains Abell370 ($3.91\times 4.45$ arcmin with $4164\times 4634$ pixels). As shown in this figure, these two detection results locate the same CGSL with a little horizontal shift which indicates the detection results of our algorithm is reliable. Original image of Abell370 by NASA, ESA, A. Koekemoer, M. Jauzac, C. Steinhardt, and the BUFFALO team.}
\label{figure18}
\end{figure}

To better investigate the performance of our algorithm. We have further designed a test. Since there is normally only one CGSL in observation images, we select the detection result with the highest score as the output. However, we could also output several detection results with high scores to test whether the detection results are reliable. Based on this concept, we output two detection results with top 2 scores of an image which contains Abell 370 in the second test, as shown in figure \ref{figure18}. In this figure, we could find that our algorithm could detect the main part of CGSL. There are only tens of pixels shift in these detection results.\\

\begin{figure*}
\centering
\subfigure[]{
\includegraphics[width=0.3\textwidth]{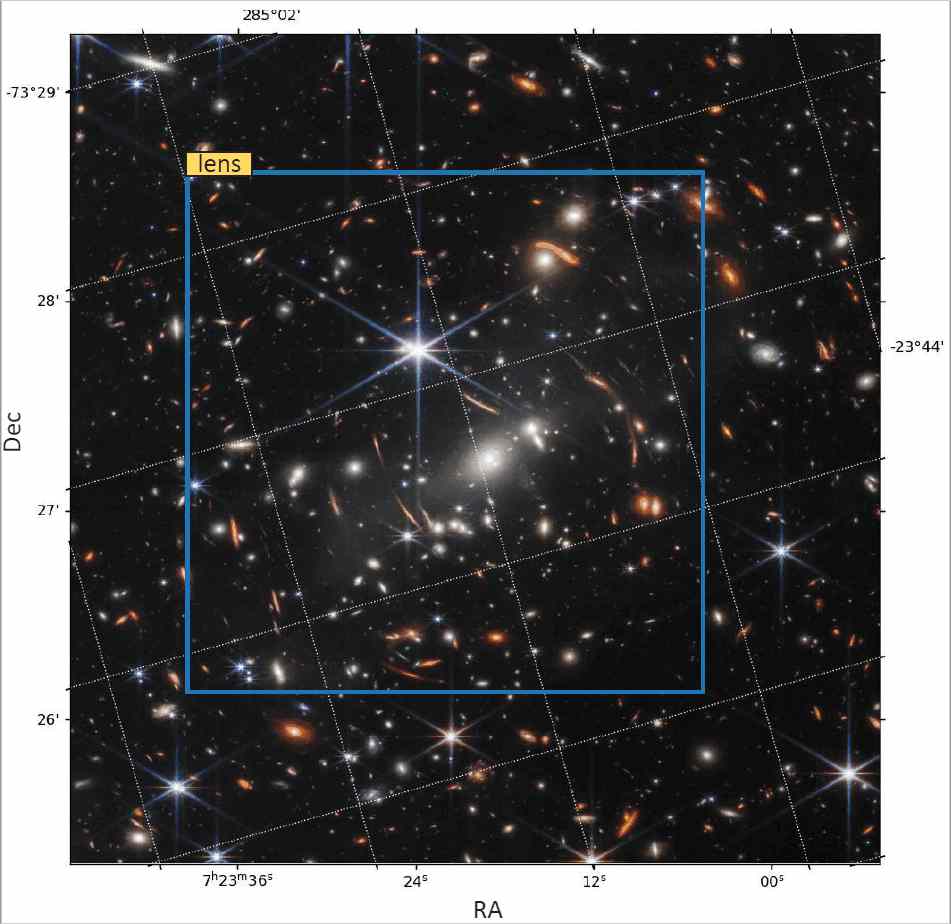}
}
\quad
\subfigure[]{
\includegraphics[width=0.3\textwidth]{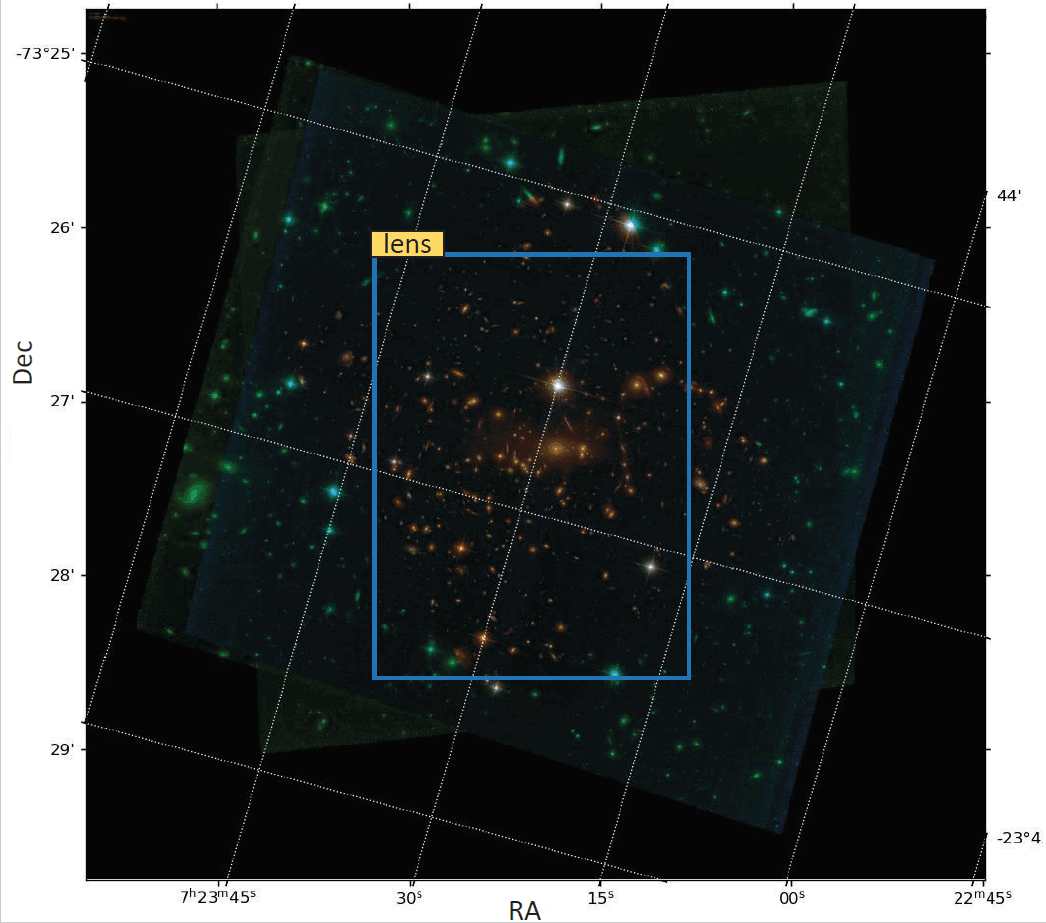}
}\\
\subfigure[]{
\includegraphics[width=0.6\textwidth]{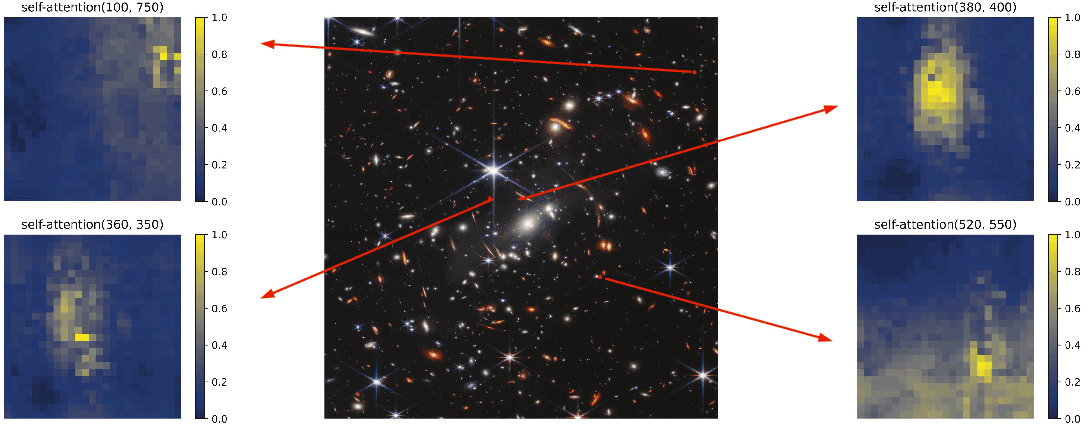}
}
\quad
\subfigure[]{
\includegraphics[width=0.27\textwidth]{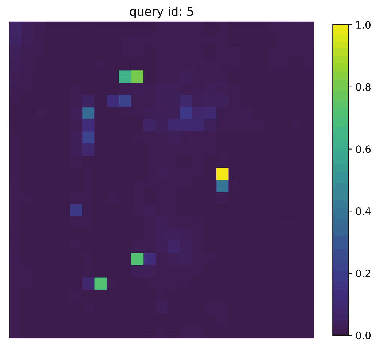}
}\\
\subfigure[]{
\includegraphics[width=0.6\textwidth]{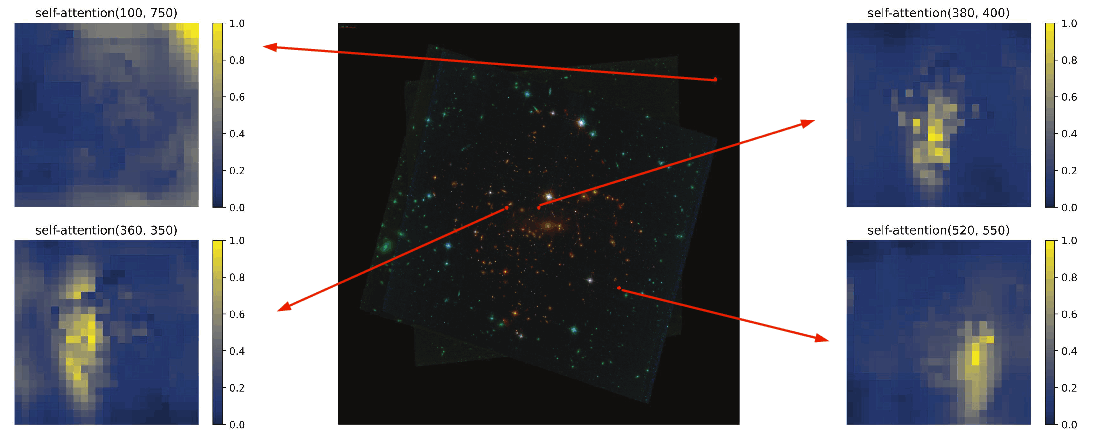}
}
\quad
\subfigure[]{
\includegraphics[width=0.27\textwidth]{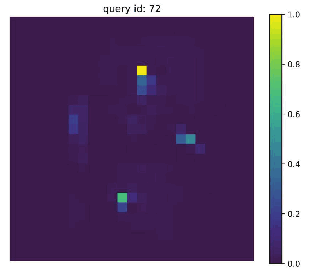}
}\\
\caption {These figures show self-attention and cross-attention of our algorithm in detection CGSLs from observation images. These two images contain the same target SMACS0723 observed by the Hubble Space Telescope and the James Webb Space Telescope. As shown in this figure, we could find that our algorithm could detect locations of the CGSLs in both of these two figures. Besides, attention maps are similar to each other, which indicates that our method could reveal similarities between these two images.}
\label{figure20}
\end{figure*}

At last, we test the performance of our algorithm in processing the same CGSL obtained by telescopes with different pixel scales. The SMACS0723 has attracted a lot of attentions recently, since the observation data by the James Webb Space Telescope was released on July 14th. 2022 \footnote{\url{https://www.nasa.gov/image-feature/goddard/2022/nasa-s-webb-delivers-deepest-infrared-image-of-universe-yet}}. The SMACS0723 has been observed by the Hubble Space Telescope in the RELICS survey project \citep{Salmon2018, Salmon2020}. In this paper, we would test the performance of our algorithm with images of the SMACS0723 observed by the Hubble Space Telescope and the James Webb Space Telescope. The image of SMACS0723 from the Hubble Space Telescope is obtained from the official site of the RELICS project as a colour image \footnote{\url{https://relics.stsci.edu/data/smacs0723-73/}}. The image of SMACS0723 from the James Webb Telescope is obtained from the official site as a colour image. Detection results and attention maps are shown in figure \ref{figure20}. As shown in this figure, our algorithm could detect and locate CGSLs in both of these two figures and attentions of these detection results are similar. According to these results, we can find that, even for real observation images, our algorithm could still obtain effective results, when trained only with simulated images. \\

\section{Conclusions and future work} \label{sec:con}
CGSLs are valuable in scientific research of galaxies, galaxy clusters, and cosmology. Due to their significant sky coverage and depth, future sky surveys are expected to reveal many CGSLs. However, considering the complexity and variousness of the strong lensing signals in CGSLs and the contamination of the line of sight objects, it is challenging to detect strong lensing signals efficiently and automatically. Hence, we propose a transformer-based algorithm for detecting CGSLs from enormous data to solve the issue. The algorithm is designed to learn features with a large extension because CGSLs are sparsely distributed and often have extended structures, which is adequate for the transformer-based detection algorithm. Our method uses ensemble learning to merge detection results from the DETR and the Deformable DETR to give final detection results. Besides, our method could detect CGSLs directly from images with any number of channels (bands), which makes it adequate in the application of detection CGSLs from multi-color sky survey projects. We use simulated images to train our algorithm, and the results show that our algorithm takes advantage of the attention mechanism and could achieve an $88\%$ recall rate and $70\%$ precision rate in detecting CGSLs, although many of them are blended with foreground galaxies. We use self-attention and cross-attention to show features that attract our detection algorithm. The results show that our algorithm focuses on the arc-like structure of CGSLs.\\

We have also considered applications of our method in real applications. To further increase detection efficiency, we propose a two-step strategy, which firstly obtain candidate images that contain CGSLs with small IOU threshold and then detect CGSLs from these candidate images with high IOU threshold. With this strategy, our method achieves a $99.63 \%$ accuracy rate, $90.32 \%$ recall rate, $85.37 \%$ precision rate and $0.23\%$ false positive rate in detecting CGSLs from 16000 images containing $1\%$ CGSLs. We further use our method to detect CGSLs from real observation images from the Hubble Space Telescope and the James Webb Space Telescope. The results show that our method can identify most of the strongly lensed arcs but miss a couple due to diffraction rings, which can be improved with more training data containing realistic PSFs and noises. Moreover, when applying our method to the HST data, we find that elongated galaxies (panel (a) in Figure~\ref{figure16}) can be the primary sources of false positives. This failure spots out two issues in the training set that need to be improved: 1) including bright stars beyond extragalactic objects; 2) making a more aggressive definition for giant arcs to eliminate the contamination due to elongated galaxies. Alternatively, we expect to improve the performance of detecting CGSLs by involving humans in the loop, which is thoroughly studied in another project of ours. \\

Compared to other machine learning-based algorithms, which try to find strong lenses by identifying stamp images centered at galaxies containing CGSLs, our algorithm focuses on recognizing arc-like features in an arbitrary field of view. Our strategy is specifically suitable for cluster-scale lenses because arcs in galaxy clusters are not guaranteed to locate around the centers of BCGs of galaxy clusters. On the other hand, compared to traditional arc-finder algorithms, our method can detect and locate the strong gravitational lensing system directly on the original image without the need to search and cut galaxy clusters and then classify them, so it is more simple and efficient. At the same time, due to the application of the attention mechanism, our algorithm has better detection performance and robustness, particularly for the cases of faint arcs, complex arcs, and highly blended images, particularly when multiple-band information is taken into account in the future. Therefore, we optimistically foresee the application of the method to the data from upcoming large-scale surveys.\\

In the future, we will carry out simulations with multiple-band information, PSFs, and noises to generate more appropriate training sets for the data obtained by different instruments. Additionally, we plan to design multi-step detection strategies and involve joint training by combining the data from various telescopes to improve the precision and recall of the detection of arcs in galaxy clusters. Eventually, the program will be applied to available and upcoming observations, such as the DESI Legacy surveys, the CSST, and the Euclid, and the data products will be released for exploring the dark sectors of the Universe.\\

\section{Acknowledgements}
This work is based on observations made with the NASA/ESA Hubble Space Telescope, obtained at the Space Telescope Science Institute, which is operated by the Association of Universities for Research in Astronomy, Inc., under NASA contract NAS 5-26555. These observations are associated with programs GO-11103, GO-12166, GO-12884, GO-14096.\\ 
This work is based on observations made with the NASA/ESA/CSA James Webb Space Telescope. The data were obtained from the Mikulski Archive for Space Telescopes at the Space Telescope Science Institute, which is operated by the Association of Universities for Research in Astronomy, Inc., under NASA contract NAS 5-03127 for JWST. These observations are associated with program \#2736.\\
Peng Jia would like to thank Professor Ran Li from National Astronomical Observatories and Professor Huanyuan Shan, Professor Shiyin Shen and Dr. Rafael S. de Souza from Shanghai Observatory who provide very helpful suggestions for this paper. This work is supported by National Natural Science Foundation of China (NSFC) with funding number of 12173027 and 12173062. We acknowledge the science research grants from the China Manned Space Project with NO. CMS-CSST-2021-A01 and CMS-CSST-2021-B12. We acknowledge the science research grants from the Square Kilometre Array (SKA) Project with NO. 2020SKA0110102. This work is also supported by the special fund for Science and Technology Innovation Teams of Shanxi Province. \\

\section*{Data Availability}
The code and data used in this paper are released in PaperData Repository powered by China-VO with a DOI number of \href{DOI: 10.12149/101172}{DOI: 10.12149/101172}.\\

\bibliography{SLT}{}
\bibliographystyle{aasjournal}



\end{document}